\documentclass[a4paper]{article}

\usepackage[utf8]{inputenc}
\usepackage[top=10mm]{geometry}
\usepackage{booktabs}
\usepackage[flushleft]{threeparttable}
\usepackage{authblk}
\usepackage{bm}
\usepackage{amsmath}
\usepackage{amssymb}
\usepackage{graphicx}
\usepackage[square,sort,numbers,compress]{natbib}
\usepackage{slashed}
\usepackage{float}
\usepackage{longtable,booktabs}
\usepackage[colorlinks,
linkcolor=blue,
anchorcolor=blue,
urlcolor=blue,
citecolor=blue
]{hyperref}
 \allowdisplaybreaks[4]

\title{\large \textbf{Phase shifts of the light pseudoscalar meson and heavy meson scattering in heavy meson chiral perturbation theory}}

\author[1]{\small Bo-Lin Huang \thanks{blhuang@pku.edu.cn}}
\author[1]{\small Zi-Yang Lin }
\author[1]{\small Kan Chen}
\author[1]{\small Shi-Lin Zhu \thanks{zhusl@pku.edu.cn}}

\affil[1]{\textit{\small School of Physics and Center of High Energy
Physics, Peking University, Beijing 100871, China}}

\date{\small \today}

\begin{document}
\maketitle

\begin{abstract}
We calculate the complete $T$ matrices of the elastic light
pseudoscalar meson and heavy meson scattering to the third order in
heavy meson chiral perturbation theory. We determine the low-energy
constants by fitting the phase shifts and scattering lengths from lattice QCD simulations simultaneously and predict the phase
shifts at the physical meson masses. The phase shifts in the
$D\pi(I=1/2)$, $DK(I=0)$, $D\bar{K}(I=0)$, $D_s\bar{K}$, $D\eta$ and
$D_s\eta$ $S$ waves are so strong that bound states or resonances
may be generated dynamically in all these channels. The $DK(I=0)$
channel corresponds to the well-known exotic state
$D_{s0}^{*}(2317)$. The $DK(I=0)$ channel corresponds
to the well-known exotic state $D_{s0}^{*}(2317)$. The
coupled-channel $D\pi$, $D\eta$ and $D_s\bar{K}$ scattering
corresponds to $D_{0}^{*}(2400)$. The coupled-channel $D\pi$, $D\eta$ and
$D_s\bar{K}$ scattering corresponds to $D_{0}^{*}(2400)$. We
also predict the scattering lengths and scattering volumes and
observe good convergence in the scattering volumes.
Our calculations provide a possibility to accurately investigate the
exotic state in the light pseudoscalar meson and heavy meson
interactions.

\begin{description}
\item[Keywords:]
Chiral perturbation theory, meson-meson scattering, phase shifts
\end{description}
\end{abstract}

\section{Introduction}

Investigations of the meson-meson scattering allow to discover interesting features of quantum chromodynamics (QCD) at hadronic energy scales, and also provide a basis for further research on hadron spectroscopy. In the past decades, the precision experimental data involving the charm quark have revealed many surprising features in hadron spectroscopy. For instance, some of the charmonium-like XYZ states lie very close to the two-meson thresholds.

As the fundamental theory of strong interaction, QCD becomes nonperturbative at low
energies. Therefore, it is very difficult to use perturbative
methods to derive the meson-meson interactions. Weinberg proposed
an effective field theory (EFT) for the purpose of solving this
problem in a seminal paper \cite{wein1979}. The EFT is formulated in
terms of the most general Lagrangian consistent with the general
symmetry principles, and the degrees of freedom are hadrons at low
energy. The corresponding formalism is called chiral
perturbation theory (ChPT) \cite{sche2012}. ChPT is a useful and
efficient tool to study hadronic physics at low energies
\cite{mach2011}. However, a power-counting problem in heavy hadron
ChPT occurs because of the nonvanishing heavy hadron mass in the
chiral limit. Heavy baryon chiral perturbation theory (HBChPT)
was proposed and developed to solve the power-counting problem that occurs in baryon ChPT \cite{gass1988,jenk1991,bern1992}. Many
achievements have been obtained in the light flavor hadronic physics
using SU(2) HBChPT
\cite{ordo1992,epel1998,fett1998, fett2000,kais19971,kang2014,ente2015,kais2020}.
Furthermore, the investigations in the SU(3) HBChPT also led to
reasonable predictions
\cite{kais2001,liu20071,haid2013,huan2015,huan2017,huan20201,huan20202,huan2021}.
The infrared regularization of the covariant baryon ChPT
\cite{bech1999} and the extended-on-mass-shell scheme
\cite{gege1999,fuch2003} for solving the power-counting problem are
two popular relativistic approaches and have led to substantial
progress in many aspects as documented in
refs.~\cite{schi2007,geng2008,alar2012,ren2012,chen2013,yao2016,lu2022}.

However, HBChPT is still a well-established and versatile tool for
the study of the low-energy hadronic physics. The expansion in HBChPT is expanded simultaneously in terms of $p/\Lambda_\chi$ and $p/M_0$, where $p$ represents the meson momentum or its mass or the small residue momentum of a baryon in the nonrelativistic limit. Similar to the HBChPT formalism in the light flavor meson-baryon and
baryon-baryon interactions, we can use the heavy meson chiral
perturbation theory (HMChPT) to address the charmed mesons, as done in ref.~\cite{wise1992}. This framework can also be extended to the heavy
flavor hadron interactions and new hadron states
\cite{wang2019,meng2020,wang2020,chen2022} (for a review of the heavy
hadron systems in ChPT, see ref.~\cite{meng2022}).

Since the discoveries of the charm-strange meson $D^{*}_{s0}(2317)$
\cite{aube2003,krok2003,bess2003} and the hidden-charm meson
$X(3872)$ \cite{choi2003}, many investigations have been devoted to
various exotic states that cannot be classified into conventional hadrons
\cite{liux2013,xie2017,ma2018,liu2019,lian2020,dong2020,wu2021,deng2022,deng20221,he2022,chenhx2022,wang2022,dai2022}.
The $D_{s0}^{*}(2317)$ has inspired  various explanations with the different methods and pictures
\cite{barn2003,beve2003,bali2003,dmit2005,guo2006,guo2007,flyn2007,lutz2008,mohl2013,liu2013,alex2020,tan2021,cheu2021,yang2022}
(for a detailed review see ref.~\cite{chen2017}). Some lattice QCD
simulations \cite{liu2013,alex2020,cheu2021} seem to support the
interpretation of the $D_{s0}^{*}(2317)$ as a $DK$ molecule. Thus,
a detailed study of the $DK$ scattering will help us to understand
the nature of this exotic state. However, in lattice calculations, the light pseudoscalar meson masses are always larger than their physical masses because of the shortage of computational resources at physical quark masses.
Therefore, the extrapolation of the light pseudoscalar meson and
heavy meson scattering from the nonphysical meson mass to the
physical value is necessary with the help of ChPT.

In our previous papers \cite{liu2009,huan2022}, we calculated
the light pseudoscalar meson and heavy meson scattering lengths up
to $\mathcal{O}(p^4)$ in HMChPT. The scattering lengths were
calculated through both perturbative and iterated methods as described in ref.~\cite{huan2022}. The value of the scattering length for the
channel $DK(I=0)$, which involves $D_{s0}^{*}(2317)$, was obtained correctly with the iterated method. In fact, the channel $DK(I=0)$ has a sufficiently strong attractive interaction and can lead to a quasi-bound state with the iterated methods, as shown in
refs.~\cite{liu2013,lutz2004,guo2009,geng2010,wang2012,alte2014,yao2015,guo2019}.
Note that a repulsive interaction has a negative scattering length
or phase shift in our convention. The scattering length is an
important quantity of the scattering process, which encodes the
information of the underlying interaction. 

The partial-wave phase shifts contain the complete information of a scattering process in the physical region. In this work, our study is concerned not only with the scattering lengths but also with the partial-wave phase shifts. We calculate the complete $T$ matrices of the elastic pseudoscalar meson and heavy meson scattering to the third order in HMChPT. Then, we determine the low-energy constants (LECs) by fitting the
phase shifts and scattering lengths simultaneously. From the complete pseudoscalar meson and heavy meson scattering amplitudes up to $\mathcal{O}(p^3)$, we can judge directly whether the attraction in a scattering channel is strong enough to generate a bound state. Furthermore, the detailed features of QCD at hadronic energy scales can be obtained from the phase shifts based on the $\mathcal{O}(p^3)$ calculation. Hopefully, the phase shifts of the light meson and heavy meson scattering (e.g., the $D^{-}K^{+}$ channel) may be extracted from the LHCb group or BelleII measurements in the future.

This paper is organized as follows. In Sec.~\ref{lagrangian}, the
chiral Lagrangians are presented up to $\mathcal{O}(p^3)$. In
Sec.~\ref{tmatrices}, the Feynman diagrams and the results of the
$T$ matrices are presented. In Sec.~\ref{phaseshifts}, we outline
how to derive partial-wave phase shifts and scattering lengths from
the $T$ matrices. Section~\ref{results} contains the numerical
results and discussions. The last section gives a brief summary.

\section{Chiral Lagrangian}
\label{lagrangian}

Our calculation of the elastic light pseudoscalar meson and heavy
meson scattering is based on the effective chiral Lagrangian in
HMChPT,
\begin{align}
\label{lagrang}
\mathcal{L}_{\text{eff}}=\mathcal{L}_{\phi\phi}+\mathcal{L}_{H
\phi}.
\end{align}
Here, the SU(3) matrix $\phi$ represents the pseudoscalar Goldstone
fields ($\phi=\pi, K, \bar{K}, \eta$). The lowest-order chiral
Lagrangian for the Goldstone meson-meson interaction takes the form
\cite{bora1997}
\begin{align}
\label{lagphiphi2} \mathcal{L}^{(2)}_{\phi\phi}=f^2\text{tr}(u_\mu
u^\mu +\frac{\chi_{+}}{4}).
\end{align}
The axial vector quantity
$u^\mu=\frac{i}{2}\{\xi^{\dagger},\partial^\mu\xi\}$ contains an
odd number of meson fields. The SU(3) matrix
$U=\xi^2=\text{exp}(i\phi/f)$ collects the pseudoscalar Goldstone
boson fields. The quantities
$\chi_{\pm}=\xi^{\dagger}\chi\xi^{\dagger} \pm \xi\chi\xi$ with
$\chi=\text{diag}(m_\pi^2,m_\pi^2,2m_K^2-m_\pi^2)$ introduce
explicit chiral symmetry breaking terms. The parameter $f$ is the
pseudoscalar decay constant in the chiral limit. The lowest-order
chiral Lagrangian for the heavy mesons in the heavy quark symmetry
limit can be written as
\begin{align}
\label{lagphiH1}
 \mathcal{L}_{H \phi}^{(1)}=-\left\langle(iv\cdot\partial H)\bar{H}\right\rangle+\left\langle H v\cdot \Gamma \bar{H}\right\rangle+g\left\langle H u_\mu \gamma^\mu \gamma_5 \bar{H}\right\rangle,
\end{align}
where $v_\mu=(1,0,0,0)$ is the heavy meson velocity, $\left\langle ... \right\rangle$ means the trace for gamma matrices, the chiral
connection $\Gamma^\mu=\frac{i}{2} [\xi^{\dagger},\partial^\mu\xi]$
contains an even number of meson fields and the doublet of the ground
state heavy mesons reads
\begin{align}
\label{Hoperater}
H=\frac{1+\slashed{v}}{2}(P_\mu^{*}\gamma^{\mu}+iP\gamma_5),\quad
\bar{H}=\gamma^{0}H^{\dag}\gamma^{0}=(P_\mu^{*\dag}\gamma^{\mu}+iP^{\dag}\gamma_5)\frac{1+\slashed{v}}{2},
\end{align}
\begin{align}
\label{Poperater} P=(D^{0},D^{+},D_s^{+}),\quad
P_\mu^{*}=(D^{0*},D^{+*},D_{s}^{+*})_\mu.
\end{align}
For the calculation of the complete $T$ matrices up to the third
order, the heavy meson Lagrangians $\mathcal{L}^{(2)}_{H \phi }$ and
$\mathcal{L}^{(3)}_{H \phi }$ in the heavy quark symmetry limit read
\begin{align}
\label{lagphiH2}
\mathcal{L}_{H \phi }^{(2)}=&c_0\left\langle H\bar{H}\right\rangle \text{tr}(\chi_{+})+c_1\left\langle H\chi_{+}\bar{H}\right\rangle-c_2\left\langle H\bar{H}\right\rangle \text{tr}(v\cdot u\,v\cdot u)-c_3\left\langle Hv\cdot u \, v\cdot u \bar{H}\right\rangle\nonumber\\
&-c_4\left\langle H\bar{H}\right\rangle \text{tr}(u^\mu
u_\mu)-c_5\left\langle H u^\mu u_\mu \bar{H}\right\rangle,
\end{align}
\begin{align}
\label{lagphiH3} \mathcal{L}_{H \phi }^{(3)}=\kappa_1 \left\langle
H[\chi_{-},v\cdot u]\bar{H}\right\rangle+i \kappa_2 \left\langle
H[v\cdot u,[v\cdot \partial,v\cdot u]]\bar{H}\right\rangle+i
\kappa_3 \left\langle H[u^\mu,[v\cdot \partial,
u_\mu]]\bar{H}\right\rangle.
\end{align}

\section{$T$ matrices}
\label{tmatrices}

\begin{figure}[t]
\centering
\includegraphics[height=10cm,width=8cm]{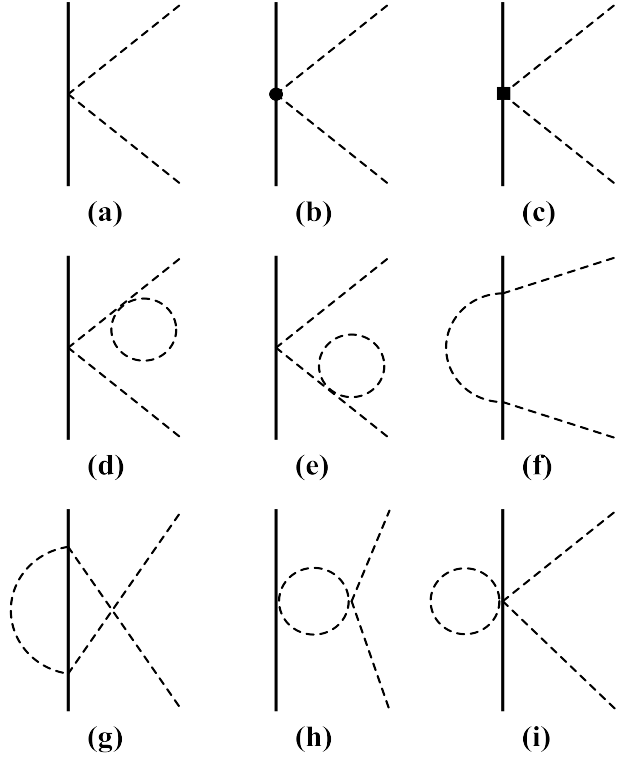}
\caption{\label{fig:feynman}Tree and nonvanishing loop diagrams in
the calculation of the Goldstonen-heavy meson scattering amplitudes to the third order in HMChPT. The dashed lines represent the Goldstone bosons and solid lines represent the pseudoscalar heavy mesons. The heavy dots and filled squares refer to the vertices from $\mathcal{L}_{ H \phi
}^{(2)}$ and $\mathcal{L}_{H \phi }^{(3)}$, respectively.}
\end{figure}

In this work, we are considering only the elastic light pseudoscalar
meson and heavy meson scattering processes
$M(\bm{q})+H(-\bm{q})\rightarrow
M(\bm{q^{\prime}})+H(-\bm{q^{\prime}})$ in the center-of-mass frame
with $|\bm{q}|=|\bm{q^{\prime}}|=q$. The leading order (LO)
amplitudes resulting from diagram (a) in Fig.~\ref{fig:feynman} read
\begin{align}
\label{TmaLO}
&T_{\pi D}^{(1/2,\text{LO})}=\frac{2w_{\pi}}{f_{\pi}^2},\quad T_{\pi D}^{(3/2,\text{LO})}=-\frac{w_{\pi}}{f_{\pi}^2},\quad T_{\pi D_{s}}^{(\text{LO})}=0,\quad T_{KD}^{(1,\text{LO})}=0,\quad T_{KD}^{(0,\text{LO})}=\frac{2w_K}{f_K^2},\quad\nonumber\\
& T_{KD_s}^{(\text{LO})}=-\frac{w_K}{f_K^2},\quad
T_{\bar{K}D}^{(1,\text{LO})}=-\frac{w_K}{f_K^2},\quad
T_{\bar{K}D}^{(0,\text{LO})}=\frac{w_K}{f_K^2},\quad
T_{\bar{K}D_s}^{(\text{LO})}=\frac{w_K}{f_K^2},\quad T_{\eta
D}^{(\text{LO})}=0,\quad T_{\eta D_s}^{(\text{LO})}=0,
\end{align}
where $K=(K^{+},K^{0})^{T}$, $\bar{K}=(\bar{K}^{0},K^{-})^{T}$. The
first superscripts of the $T$ matrices denote the total isospin. In
the channels with an isoscalar $\eta$-meson or $D_s$-meson, the total
isospin is unique and does not need to be specified. The quantities
$w_{\phi}=(m_{\phi}^2+q^2)^{1/2}$ with $\phi=(\pi,K,\eta)$ denote
the center-of-mass energy of the light pseudoscalar mesons. We take
the renormalized decay constants $f_{\phi}$ with the nonzero quark mass
instead of $f$ in the chiral limit.

At the next-to-leading order (NLO), one has the contributions from diagram (b) of Fig.~\ref{fig:feynman}, which involves the vertex
from the Lagrangian $\mathcal{L}_{H \phi }^{(2)}$. The amplitudes
involving the low-energy constants (LECs) read
\begin{align}
\label{TmaNLOpiD1c2} T_{\pi
D}^{(1/2,\text{NLO})}=&\frac{1}{f_\pi^2}[8c_0m_\pi^2+4c_1m_\pi^2+2c_2w_\pi^2+c_3w_\pi^2+2c_4(w_\pi^2-q^2
z)+c_5(w_\pi^2-q^2 z)],
\end{align}
\begin{align}
\label{TmaNLOpiD3c2} T_{\pi
D}^{(3/2,\text{NLO})}=&\frac{1}{f_\pi^2}[8c_0m_\pi^2+4c_1m_\pi^2+2c_2w_\pi^2+c_3w_\pi^2+2c_4(w_\pi^2-q^2
z)+c_5(w_\pi^2-q^2 z)],
\end{align}
\begin{align}
\label{TmaNLOpiDs} T_{\pi
D_s}^{(\text{NLO})}=\frac{1}{f_\pi^2}[8c_0m_\pi^2+2c_2w_\pi^2+2c_4(w_\pi^2-q^2
z)],
\end{align}
\begin{align}
\label{TmaNLOKD1} T_{K
D}^{(1,\text{NLO})}=\frac{1}{f_K^2}[8c_0m_K^2+2c_2w_K^2+2c_4(w_K^2-q^2z)],
\end{align}
\begin{align}
\label{TmaNLOKD0} T_{K
D}^{(0,\text{NLO})}=&\frac{1}{f_K^2}[8c_0m_K^2+8c_1m_K^2+2c_2w_K^2+2c_3w_K^2+2c_4(w_K^2-q^2
z)+2c_5(w_K^2-q^2 z)],
\end{align}
\begin{align}
\label{TmaNLOKDs} T_{K
D_s}^{(\text{NLO})}=&\frac{1}{f_K^2}[8c_0m_K^2+4c_1m_K^2+2c_2w_K^2+c_3w_K^2+2c_4(w_K^2-q^2
z)+c_5(w_K^2-q^2 z)],
\end{align}
\begin{align}
\label{TmaNLOKbarD1} T_{\bar{K}
D}^{(1,\text{NLO})}=&\frac{1}{f_K^2}[8c_0m_K^2+4c_1m_K^2+2c_2w_K^2+c_3w_K^2+2c_4(w_K^2-q^2
z)+c_5(w_K^2-q^2 z)],
\end{align}
\begin{align}
\label{TmaNLOKbarD0} T_{\bar{K}
D}^{(0,\text{NLO})}=&\frac{1}{f_K^2}[8c_0m_K^2-4c_1m_K^2+2c_2w_K^2-c_3w_K^2+2c_4(w_K^2-q^2
z)-c_5(w_K^2-q^2 z)],
\end{align}
\begin{align}
\label{TmaNLOKbarDs} T_{\bar{K}
D_s}^{(\text{NLO})}=&\frac{1}{f_K^2}[8c_0m_K^2+4c_1m_K^2+2c_2w_K^2+c_3w_K^2+2c_4(w_K^2-q^2
z)+c_5(w_K^2-q^2 z)],
\end{align}
\begin{align}
\label{TmaNLOetaD} T_{\eta
D}^{(\text{NLO})}=\frac{1}{3f_\eta^2}[24c_0
m_\eta^2+4c_1m_\pi^2+6c_2w_\eta^2+c_3w_\eta^2+6c_4(w_\eta^2-q^2
z)+c_5(w_\eta^2-q^2 z)],
\end{align}
\begin{align}
\label{TmaNLOetaDs} T_{\eta
D_s}^{(\text{NLO})}=&\frac{1}{3f_\eta^2}[24c_0
m_\eta^2+16c_1(2m_K^2-m_\pi^2)+6c_2w_\eta^2+4c_3w_\eta^2+6c_4(w_\eta^2-q^2
z)+4c_5(w_\eta^2-q^2 z)],
\end{align}
where $z=\text{cos}\,\theta$ is the cosine of the angle $\theta$
between $\bm{q}$ and $\bm{q^{\prime}}$.

At the next-to-next-to-leading order (N2LO), one has contributions
from diagram (c) in Fig.~\ref{fig:feynman}, which involves the
vertex from the Lagrangian $\mathcal{L}_{H \phi }^{(3)}$. The
amplitudes read
\begin{align}
\label{TmaN2LOpiD1c2} T_{\pi
D}^{(1/2,\text{N2LO})}=\frac{1}{f_\pi^2}[16\bar{\kappa}_1
m_\pi^2w_\pi+4\bar{\kappa}_2 w_\pi^{3}+4\bar{\kappa}_3 w_\pi
(w_\pi^2-q^2z)],
\end{align}
\begin{align}
\label{TmaN2LOpiD3c2} T_{\pi
D}^{(3/2,\text{N2LO})}=\frac{1}{f_\pi^2}[-8\bar{\kappa}_1 m_\pi^2
w_\pi-2\bar{\kappa}_2 w_\pi^{3}-2\bar{\kappa}_3w_\pi(w_\pi^2-q^2
z)],
\end{align}
\begin{align}
\label{TmaN2LOpiDs} T_{\pi D_s}^{(\text{N2LO})}=0,
\end{align}
\begin{align}
\label{TmaN2LOKD1} T_{K D}^{(1,\text{N2LO})}=0,
\end{align}
\begin{align}
\label{TmaN2LOKD0} T_{K
D}^{(0,\text{N2LO})}=\frac{1}{f_K^2}[16\bar{\kappa}_1
m_K^2w_K+4\bar{\kappa}_2 w_K^{3}+4\bar{\kappa}_3w_K(w_K^2-q^2 z)],
\end{align}
\begin{align}
\label{TmaN2LOKDs} T_{K
D_s}^{(\text{N2LO})}=\frac{1}{f_K^2}[-8\bar{\kappa}_1 m_K^2
w_K-2\bar{\kappa}_2 w_K^{3}-2\bar{\kappa}_3 w_K (w_K^2-q^2 z)],
\end{align}
\begin{align}
\label{TmaN2LOKbarD1} T_{\bar{K}
D}^{(1,\text{N2LO})}=\frac{1}{f_K^2}[-8\bar{\kappa}_1 m_K^2
w_K-2\bar{\kappa}_2 w_K^{3}-2\bar{\kappa}_3 w_K (w_K^2-q^2 z)],
\end{align}
\begin{align}
\label{TmaN2LOKbarD0} T_{\bar{K}
D}^{(0,\text{N2LO})}=\frac{1}{f_K^2}[8\bar{\kappa}_1 m_K^2
w_K+2\bar{\kappa}_2 w_K^{3}+2\bar{\kappa}_3 w_K (w_K^2-q^2 z)],
\end{align}
\begin{align}
\label{TmaN2LOKbarDs} T_{\bar{K}
D_s}^{(\text{N2LO})}=\frac{1}{f_K^2}[8\bar{\kappa}_1 m_K^2
w_K+2\bar{\kappa}_2 w_K^{3}+2\bar{\kappa}_3 w_K (w_K^2-q^2 z)],
\end{align}
\begin{align}
\label{TmaN2LOetaD} T_{\eta D}^{(\text{N2LO})}=0,
\end{align}
\begin{align}
\label{TmaN2LOetaDs} T_{\eta D_s}^{(\text{N2LO})}=0.
\end{align}
At this order, one also has the amplitudes from the one-loop
diagrams. The nonvanishing one-loop diagrams generated by the
vertices of $\mathcal{L}^{(2)}_{\phi\phi}$ and $\mathcal{L}_{H
\phi}^{(1)}$ are shown in the second and third row of
Fig.~\ref{fig:feynman}. Note that the third-order scale-independent
LECs $\bar{\kappa}_{1}$, $\bar{\kappa}_{2}$ and $\bar{\kappa}_{3}$
are used in the counterterm $T$ matrices, as done in ref.~\cite{fett1998}. Putting all amplitudes from the different one-loop diagrams together, we have
\begin{align}
\label{TmaLOOPpiD1c2} T_{\pi
D}^{(1/2,\text{LOOP})}=&-\frac{w_\pi}{12f_\pi^4}\{3w_\pi[3J_0(-w_\pi,m_K)+4J_0(-w_\pi,m_\pi)-9J_0(w_\pi,m_K)-17J_0(w_\pi,m_\pi)]\nonumber\\&+12I_2(t,m_K)+16I_2(t,m_\pi)\},
\end{align}
\begin{align}
\label{TmaLOOPpiD3c2} T_{\pi
D}^{(3/2,\text{LOOP})}=&\frac{w_\pi}{6f_\pi^4}\{3w_\pi[3J_0(-w_\pi,m_K)+7J_0(-w_\pi,m_\pi)+J_0(w_\pi,m_\pi)]+3I_2(t,m_K)\nonumber\\&+4I_2(t,m_\pi)\},
\end{align}
\begin{align}
\label{TmaLOOPpiDs} T_{\pi
D_s}^{(\text{LOOP})}=&\frac{w_\pi^2}{2f_\pi^4}[J_0(w_\pi,m_K)+J_0(-w_\pi,m_K)],
\end{align}
\begin{align}
\label{TmaLOOPKD1} T_{K
D}^{(1,\text{LOOP})}=\frac{w_K}{2f_K^4}\{w_K[J_0(-w_K,m_K)+J_0(w_K,m_\pi)]+2I_2(t,m_\pi)-I_2(t,m_K)\},
\end{align}
\begin{align}
\label{TmaLOOPKD0} T_{K
D}^{(0,\text{LOOP})}=&\frac{w_K}{12f_K^4}\{3w_K[2J_0(-w_K,m_K)+24J_0(w_K,m_\eta)+22J_0(w_K,m_K)+J_0(w_K,m_\pi)]\nonumber\\&-10I_2(t,m_K)-36I_2(t,m_\pi)\},
\end{align}
\begin{align}
\label{TmaLOOPKDs} T_{K
D_s}^{(\text{LOOP})}=&\frac{w_K}{12f_K^4}\{3w_K[12J_0(-w_K,m_\eta)+7J_0(-w_K,m_K)+5J_0(-w_K,m_\pi)+2J_0(w_K,m_K)]\nonumber\\&+14I_2(t,m_K)\},
\end{align}
\begin{align}
\label{TmaLOOPKbarD1} T_{\bar{K}
D}^{(1,\text{LOOP})}=&\frac{w_K}{24f_K^4}\{3w_K[24J_0(-w_K,m_\eta)+22J_0(-w_K,m_K)+3J_0(-w_K,m_\pi)+4J_0(w_K,m_K)]\nonumber\\&+16I_2(t,m_K)+24I_2(t,m_\pi)\},
\end{align}
\begin{align}
\label{TmaLOOPKbarD0} T_{\bar{K}
D}^{(0,\text{LOOP})}=&\frac{w_K}{24f_K^4}\{3w_K[-24J_0(-w_K,m_\eta)-22J_0(-w_K,m_K)+5J_0(-w_K,m_\pi)+4J_0(w_K,m_K)]\nonumber\\&+8I_2(t,m_K)-72I_2(t,m_\pi)\},
\end{align}
\begin{align}
\label{TmaLOOPKbarDs} T_{\bar{K}
D_s}^{(\text{LOOP})}=&\frac{w_K}{12f_K^4}\{3w_K[2J_0(-w_K,m_K)+12J_0(w_K,m_\eta)+7J_0(w_K,m_K)+5J_0(w_K,m_\pi)]\nonumber\\&-14I_2(t,m_K)\},
\end{align}
\begin{align}
\label{TmaLOOPetaD} T_{\eta
D}^{(\text{LOOP})}=&\frac{3w_\eta^2}{4f_\eta^4}[J_0(-w_\eta,m_K)+J_0(w_\eta,m_K)],
\end{align}
\begin{align}
\label{TmaN2LOetaDs} T_{\eta
D_s}^{(\text{LOOP})}=&\frac{21w_\eta^2}{8f_\eta^4}[J_0(-w_\eta,m_K)+J_0(w_\eta,m_K)],
\end{align}
with the finite parts of loop functions
\begin{align}
\label{J0func} J_0(w,m) =\frac{w}{8\pi^2} +\begin{cases}
\dfrac{1}{4\pi^2}\sqrt{w^2-m^2}\text{ln}\dfrac{-w+\sqrt{w^2-m^2}}{m}& (w<-m),\\
-\dfrac{1}{4\pi^2}\sqrt{m^2-w^2}\text{arccos}\dfrac{-w}{m}& (-m<w<m),\\
\dfrac{1}{4\pi^2}\sqrt{w^2-m^2}\Bigg(i\pi-\text{ln}\dfrac{w+\sqrt{w^2-m^2}}{m}\Bigg)&(w>m),
\end{cases}
\end{align}
\begin{align}
\label{I2func}
I_2(t,m)=\frac{1}{48\pi^2}\Bigg\{2m^2-\frac{5t}{12}-\frac{(4m^2-t)^{3/2}}{2\sqrt{-t}}\text{ln}\frac{\sqrt{4m^2-t}+\sqrt{-t}}{2m}\Bigg\},
\end{align}
and the squared invariant momentum transfer $t=2q^2(z-1)$.

\section{Partial-wave phase shifts and scattering lengths}
\label{phaseshifts}

The partial-wave amplitudes $f_{l}^{(I)}(q)$, where $l$ refers to
the orbital angular momentum, are obtained from the $T$ matrix by a
projection:
\begin{align}
\label{fampfunc}
f_{l}^{(I)}(q)=\frac{M_{H}}{16\pi\sqrt{s}}\int^{+1}_{-1}dz\, [T_{\phi
H}^{(I)}P_{l}(z)],
\end{align}
where $P_{l}(z)$ denotes the conventional Legendre polynomial and
$\sqrt{s}=(m_\phi^2+q^2)^{1/2}+(M_H^2+q^2)^{1/2}$ is the total
center-of-mass energy. For the channels that may generate the bound
states or resonances, the $T$ matrix must be iterated to the infinite order. We consider the $T$ matrix up to the third order only in the calculation of the phase shifts and scattering lengths since we do not aim to achieve the description of the bound states or resonances. For the energy range considered in this paper, the phase shifts without the effect of the bound states or resonances $\delta_{l}^{(I)}(q)$ are calculated by (also see
refs.~\cite{gass1991,fett1998})
\begin{align}
\label{delta}
\delta_{l}^{(I)}(q)=\text{arctan}[q\text{Re}f_{l}^{(I)}(q)].
\end{align}
Based on relativistic kinematics, there is a relation between the
center-of-mass momentum and the momentum of the incident light
pseudoscalar meson in the laboratory system,
\begin{align}
\label{qmomentum} q^2=\frac{M_H^2
p_\text{lab}^2}{m_\phi^2+M_{H}^2+2M_H\sqrt{m_\phi^2+p_{\text{lab}}^2}}.
\end{align}
The scattering lengths for the $S$ waves and the scattering volumes
for $P$ waves are obtained by dividing out the threshold behavior of
the respective partial-wave amplitude and approaching the threshold
\cite{eric1988}
\begin{align}
\label{thresholdpar} a_{l}^{(I)}=\lim\limits_{q \rightarrow
0}q^{-2l}f_{l}^{(I)}(q).
\end{align}

\section{Results and discussion}
\label{results}

In order to determine the low-energy constants, we start by fitting
both phase shifts and scattering lengths from lattice QCD
simulations at the nonphysical meson values simultaneously, and we
then make predictions for the phase shifts and the threshold
parameters in all channels at the physical meson values.

\subsection{Fitting}

Now, we determine $c_{0,...,5}$ and $\bar{\kappa}_{1,2,3}$ using the
phase shifts and the scattering lengths from lattice data. We take
the $S$-wave phase shifts with $I=3/2$ and the $P$-wave phase shifts
with $I=1/2$ of the elastic $D\pi$ scattering at $m_\pi\simeq
391\,\text{MeV}$ from ref.~\cite{moir2016}. The $S$-wave phase shift
with $I=1/2$ of the $D\pi$ scattering is not used in the fitting
because there exists a near-threshold bound state that cannot be
obtained in the perturbative method. The $P$-wave phase shift with
$I=0$ of the elastic $DK$ scattering at $m_\pi\simeq
239\,\text{MeV}$ is taken from ref.~\cite{cheu2021}. Again, there
exists a bound state, i.e., $D_{s0}^{*}(2317)$, in the $S$-wave
$I=0$ $DK$ channel, and then the phase shift from this channel is
not used to determine the LECs. The phase shifts of the elastic
$D\bar{K}$ scattering are obtained by using a simple parametrization
with the scattering lengths in ref.~\cite{cheu2021}. Therefore, we
use the $D\bar{K}$ scattering lengths directly instead of the phase
shifts. For the three phase shifts that are used to determine the
LECs, we take the data with the pion (kaon) laboratory momentum between 5
and 300 MeV. In addition, the scattering lengths of the five
channels [$D\bar{K}(I=0)$, $D\bar{K}(I=1)$, $D\pi(I=3/2)$, $D_sK$,
$D_s\pi$] are used to determine the LECs from
refs.~\cite{moir2016,cheu2021,liu2013}. We take the scattering
length of the channel [$D\pi(I=3/2)$] at $m_\pi \simeq
391\,\text{MeV}$ from ref.~\cite{moir2016}, the scattering lengths
of the channels [$D\bar{K}(I=0)$, $D\bar{K}(I=1)$] at $m_\pi =
239\,\text{MeV}$ and $m_\pi = 391\,\text{MeV}$ from
ref.~\cite{cheu2021}, and the (M007, M010) data for the five
channels from ref.~\cite{liu2013}. The corresponding lattice values
of $f_\pi$ and $f_K$ are from ref.~\cite{walk2009}, and we always
choose $f_\eta=1.2f_\pi$ in this paper. The resulting LECs with the
correlations between the parameters can be found in
Table~\ref{fittingresult}. The uncertainty for the respective
parameter is statistical, and it measures how much a particular
parameter can be changed while maintaining a good description of the
fitting data. Nevertheless, the parameters cannot truly vary
independently of each other because of the mutual correlations, as
detailed in refs.~\cite{doba2014,carl2016}. Therefore, the large
uncertainties of some LECs in our fit cannot make the errors of the
phase shifts and the threshold parameters large because a
full error analysis requires a complete covariance matrix. However,
we obtain small uncertainties for some LECs (e.g., $c_4$, $c_5$,
$\bar{\kappa}_1$). Furthermore, the values of the LECs are mostly of
natural size; i.e., they are numbers of order one and the same order of magnitude as the axial vector coupling constant $g_A=1.27$ \cite{chan2018, mark2019}. We can see that the absolute values for most of the LECs turn out to be between one and ten when one introduces dimensionless LECs (e.g., $c_i^{'}=\Lambda_\chi c_i$). In fact, the values of the LECs in the calculations of the pion-nucleon scattering were obtained at the same order of magnitude and regarded as of natural size in refs.~\cite{fett1998,fett2000,huan20201,huan20202}. However, $c_{2,3,4}$ are around $5-9\,\text{GeV}^{-1}$, which may be enhanced by including $D_{s0}^{*}(2317)$ explicitly. In comparison, the $\Delta(1232)$ resonance enhanced the LECs in the pion-nucleon scattering. For the channel $DK(I=0)$ involving $D_{s0}^{*}(2317)$, we clearly see that $c_2$ and $c_3$ can be combined into the linear combination $c_2+c_3$, which has a small value of $0.43\,\text{GeV}^{-1}$. The absolute value of the correlation between $c_0$ and $c_1$ is very close to one, which is consistent with the fact that the terms with these two parameters involve only the masses of the light pseudoscalar mesons.

\begin{table*}[!t]
\centering
\resizebox{\textwidth}{!}{%
\begin{threeparttable}
\caption{\label{fittingresult}Results of fitting to various lattice
data of the phase shifts and scattering lengths. For a detailed
description, see the main text.}
\begin{tabular}{ccccccccccccccccccc}
\midrule \toprule
 & Values & $c_0$ & $c_1$ & $c_2$ & $c_3$ & $c_4$ & $c_5$ & $\bar{\kappa}_1$ & $\bar{\kappa}_2$ & $\bar{\kappa}_3$ & \\
\midrule
$c_0$ ($\text{GeV}^{-1}$)&$-0.77\pm 0.39$&$1.00$ &$0.99$ & $-0.86$& $-0.88$ & $0.00$ &$0.00$&$0.02$& $-0.92$ & $0.00$ &\\
\midrule
$c_1$ ($\text{GeV}^{-1}$)&$-0.64\pm 0.35$ & &$1.00$ &$-0.86$ & $-0.89$& $0.00$ & $0.00$ & $0.01$ &$-0.93$& $0.00$ &\\
\midrule
$c_2$ ($\text{GeV}^{-1}$)&$-5.04\pm 1.83$& & & $1.00$ &$0.95$ & $-0.51$& $-0.41$ & $-0.01$ &$0.61$&$0.50$&\\
\midrule
$c_3$ ($\text{GeV}^{-1}$)&$5.47\pm 1.57$&  & & & $1.00$&$-0.36$& $-0.45$ &$0.10$ &$0.67$ &$0.40$\\
\midrule
$c_4$ ($\text{GeV}^{-1}$)&$8.99\pm 0.93$&  &   & & & $1.00$&$0.80$&$0.00$&$0.37$&$-0.98$\\
\midrule
$c_5$ ($\text{GeV}^{-1}$)&$-3.08\pm 0.70$&    &    & &  & &$1.00$&$0.00$&$0.34$&$-0.90$&\\
\midrule
$\bar{\kappa}_1$ ($\text{GeV}^{-2}$)&$0.21\pm 0.04$&    &  & & & & & $1.00$& $0.01$& $0.00$&\\
\midrule
$\bar{\kappa}_2$ ($\text{GeV}^{-2}$)&$7.81\pm 3.88$&    &  & & & & & & $1.00$& $-0.38$&\\
\midrule
$\bar{\kappa}_3$ ($\text{GeV}^{-2}$)&$-1.87\pm 1.47$&    &  & &   & & & & & $1.00$&\\
\midrule
$\chi^2/\text{d.o.f.}$&$\frac{34.85}{195-9}=0.19$&    &    &\\
\bottomrule \midrule
\end{tabular}
\end{threeparttable}}%
\end{table*}

\begin{figure}[!t]
\centering
\includegraphics[height=16.5cm,width=12.5cm]{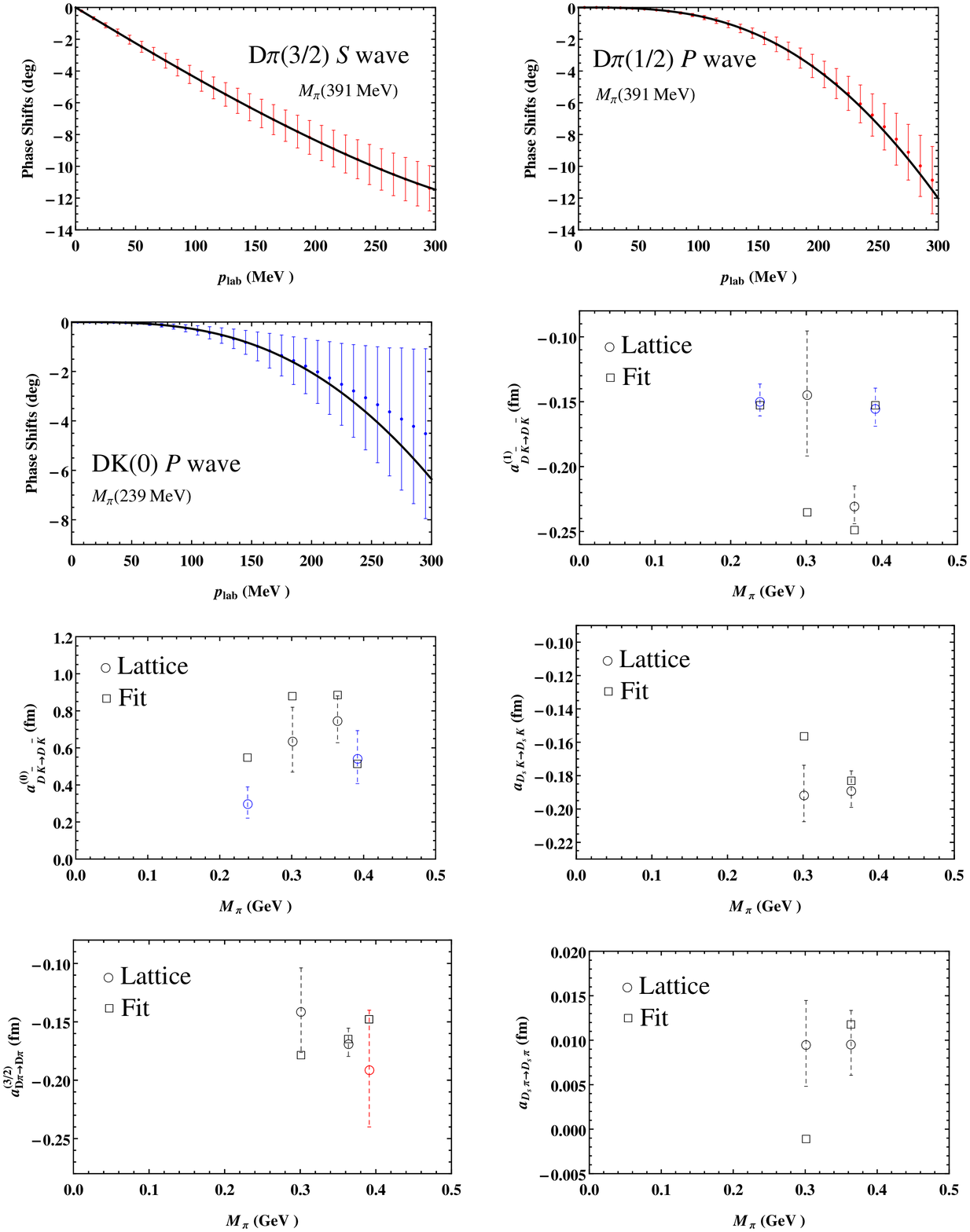}
\caption{\label{fig:fitting}Fits for the light pseudoscalar meson
and $D$ meson phase shifts and scattering lengths from various
lattice data. The lattice phase shifts and scattering length with
the red error bars in the $D\pi(I=3/2)$ $S$-wave, $D\pi(I=1/2)$
$P$-wave and $a_{D\pi\rightarrow D\pi}^{(3/2)}$ are from
ref.~\cite{moir2016}. The lattice data with the blue error bars in
the $DK(I=0)$ $P$-wave, $a_{D\bar{K}\rightarrow D\bar{K}}^{(1)}$ and
$a_{D\bar{K}\rightarrow D\bar{K}}^{(0)}$ are from
ref.~\cite{cheu2021}. The lattice scattering lengths with the black
error bars are from ref.~\cite{liu2013}. For a detailed description
of the fits, see the main text.}.
\end{figure}

The corresponding phase shifts and scattering lengths from the
fitting are shown in Fig.~\ref{fig:fitting}. The $S$-wave phase
shifts with $I=3/2$ and $P$-wave phase shifts with $I=1/2$ of the
$D\pi$ scattering at $m_\pi\simeq391\,\text{MeV}$ are in very good
agreement with the data from lattice QCD simulations up to the pion
laboratory momentum of 300 MeV. For the isoscalar $P$-wave $DK$ scattering at $m_\pi\simeq 239\,\text{MeV}$, the values of the phase shifts are very consistent with the data from lattice QCD simulations below the kaon laboratory momentum of 200 MeV. However, the $P$-wave phase shifts of $DK(I=0)$ from lattice QCD simulations have large errors. The values of the $DK(I=0)$ $P$-wave phase shifts are in agreement with the results from lattice QCD within errors up to the kaon laboratory momentum of 300 MeV. The scattering length of $D\pi(I=3/2)$ at $m_\pi\simeq 391\,\text{MeV}$ is in agreement with the lattice QCD value from ref.~\cite{moir2016} within error. The scattering lengths of $D\bar{K}(I=1)$ at $m_\pi=239,391\,\text{MeV}$ and $D\bar{K}(I=0)$ at $m_\pi=391\,\text{MeV}$ are in good agreement with the values of lattice QCD from ref.~\cite{cheu2021}. The value for $D\bar{K}(I=0)$ at $m_\pi=239\,\text{MeV}$ has a small deviation from lattice QCD. The reason is that there may exist a virtual bound state in this channel. The scattering lengths of the five channels [$D\bar{K}(I=0)$, $D\bar{K}(I=1)$, $D\pi(I=3/2)$, $D_sK$, $D_s\pi$] at $m_\pi\simeq 301,364\,\text{MeV}$ are in agreement with lattice QCD values from ref.~\cite{liu2013} within errors. There exist small deviations at a few points because the lattice QCD values are from different groups, which may cause some errors in this fitting. We have obtained a good description of the three phase shifts and the five scattering lengths at the nonphysical meson values.

Due to the strong correlations in some parameters, we use the linear combinations of the low-energy constants for further analysis. The combinations $c_2+c_4$ and $c_3+c_5$ contribute to the $S$-wave scattering. We also use $c_0+c_1$ because they have a large correlation. Thus, we have three linear combinations instead of the separate $c_i$. The results can be found in Table~\ref{fittingresultcombi}. Unsurprisingly, we obtain small values and uncertainties for the three linear combinations. However, we have omitted the difference in the description of the $P$ waves in this fitting with the linear parameter combinations. On the other hand, the values in Table~\ref{fittingresult} can describe exactly the corresponding phase shifts and threshold parameters with the help of the mutual correlations. Therefore, it is not necessary to further analyze the phase shifts and threshold parameters in this fitting. However, we can study the reason why the values of the LECs in Table~\ref{fittingresult} are larger than the three linear combinations from Table~\ref{fittingresultcombi}. It is easy to find that the LECs in $P$ waves are only $c_4$, $c_5$, and $\bar{\kappa}_3$. Thus, we fit the three LECs by using the $P$-wave phase shifts of the $D\pi(I=1/2)$ and $DK(I=0)$ channels. We obtain $c_4=13.41\pm0.22\,\text{GeV}^{-1}$, $c_5=-0.43\pm0.16\,\text{GeV}^{-1}$, and $\bar{\kappa}_3=-8.67\pm0.34\,\text{GeV}^{-2}$ with a very small $\chi^2/\text{d.o.f}= 0.01$, which is caused by the large errors in the $P$-wave phase shifts. Nevertheless, the large value for $c_4$ is also not of natural size. This may be one reason why the values in Table~\ref{fittingresult} are large. We need more precise data of the $P$ waves to improve the LECs in Table~\ref{fittingresult}.

For the channel $DK(I=0)$, we can explicitly include $D_{s0}^{*}(2317)$ in the fitting to improve the values of LECs. We note that the $\Lambda(1405)$ was included for the $KN$ scattering \cite{lee1994} and $\Delta(1232)$ was included for the $\pi N$ scattering \cite{chen2013}. Unfortunately, the coupling constant involving $D_{s0}^{*}(2317)$ has not been determined. Thus, we cannot obtain an additional constraint to determine the LECs. However, we can use the scattering length of the channel $DK(I=0)$ from lattice QCD to determine the coupling constant with $D_{s0}^{*}(2317)$. The value of the scattering length for the channel $DK(I=0)$ is $-1.33(20)\,\text{fm}$ from ref.~\cite{mohl2013} in a near threshold lattice simulation. The corresponding formula can be found in Appendix~\ref{D2317contr}. We can obtain $g_R^2=0.81\pm0.04$ by using the lattice values from ref.~\cite{mohl2013} and the mass of the $D_{s0}^{*}(2317)$ from PDG \cite{pdg2020}. The values of $g_R$ for $D_{s0}^{*}(2317)$ are not very large and have the same order of magnitude as the coupling constant involving $\Lambda(1405)$ ($\bar{g}_{\Lambda_R}^2=0.15$). Therefore, the inclusion of $D_{s0}^{*}(2317)$ in channel $DK(I=0)$ is reasonable. Global fitting can be performed after the coupling constant involving $D_{s0}^{*}(2317)$ is determined. We can see that $D_{s0}(2317)$ affects only the $S$-wave behavior of the $DK(I=0)$ channel in this method. The $P$-wave behavior should not be affected by $D_{s0}^{*}(2317)$ because $D_{s0}^{*}(2317)$ is always interpreted as a $DK$ molecule with $I(J^P)=0(0^+)$. However, the $P$-wave behavior can be improved by including the vector heavy mesons. We will discuss the issue in a forthcoming calculation.

\begin{table*}[!t]
\centering
\resizebox{\textwidth}{!}{%
\begin{threeparttable}
\caption{\label{fittingresultcombi}Results of fitting by using the linear combinations of the low-energy constants. For a detailed description, see the main text.}
\begin{tabular}{ccccccccccccccccccc}
\midrule \toprule
 & Values & $c_0+c_1$ & $c_2+c_4$ & $c_3+c_5$ & $\bar{\kappa}_1$ & $\bar{\kappa}_2$ & $\bar{\kappa}_3$ & \\
\midrule
$c_0+c_1$ ($\text{GeV}^{-1}$)&$0.04\pm 0.02$&$1.00$ &$0.01$ & $-0.47$& $-0.22$ & $-0.79$ &$0.00$&\\
\midrule
$c_2+c_4$ ($\text{GeV}^{-1}$)&$0.89\pm 0.02$& & $1.00$ &$-0.09$ & $0.01$& $0.01$ & $0.00$ &\\
\midrule
$c_3+c_5$ ($\text{GeV}^{-1}$)&$-0.32\pm 0.04$&  & & $1.00$&$0.95$& $0.77$ &$0.00$ &\\
\midrule
$\bar{\kappa}_1$ ($\text{GeV}^{-2}$)&$0.21\pm 0.07$&    & & & $1.00$& $0.60$& $0.00$&\\
\midrule
$\bar{\kappa}_2$ ($\text{GeV}^{-2}$)&$-7.13\pm 0.34$&  & & & & $1.00$& $-0.39$&\\
\midrule
$\bar{\kappa}_3$ ($\text{GeV}^{-2}$)&$6.11\pm 0.14$&    & & & & & $1.00$&\\
\midrule
$\chi^2/\text{d.o.f.}$&$\frac{147}{195-6}=0.78$&    &    &\\
\bottomrule \midrule
\end{tabular}
\end{threeparttable}}%
\end{table*}

\subsection{Phase shifts}

\begin{figure}[!t]
\centering
\includegraphics[height=12.5cm,width=12.5cm]{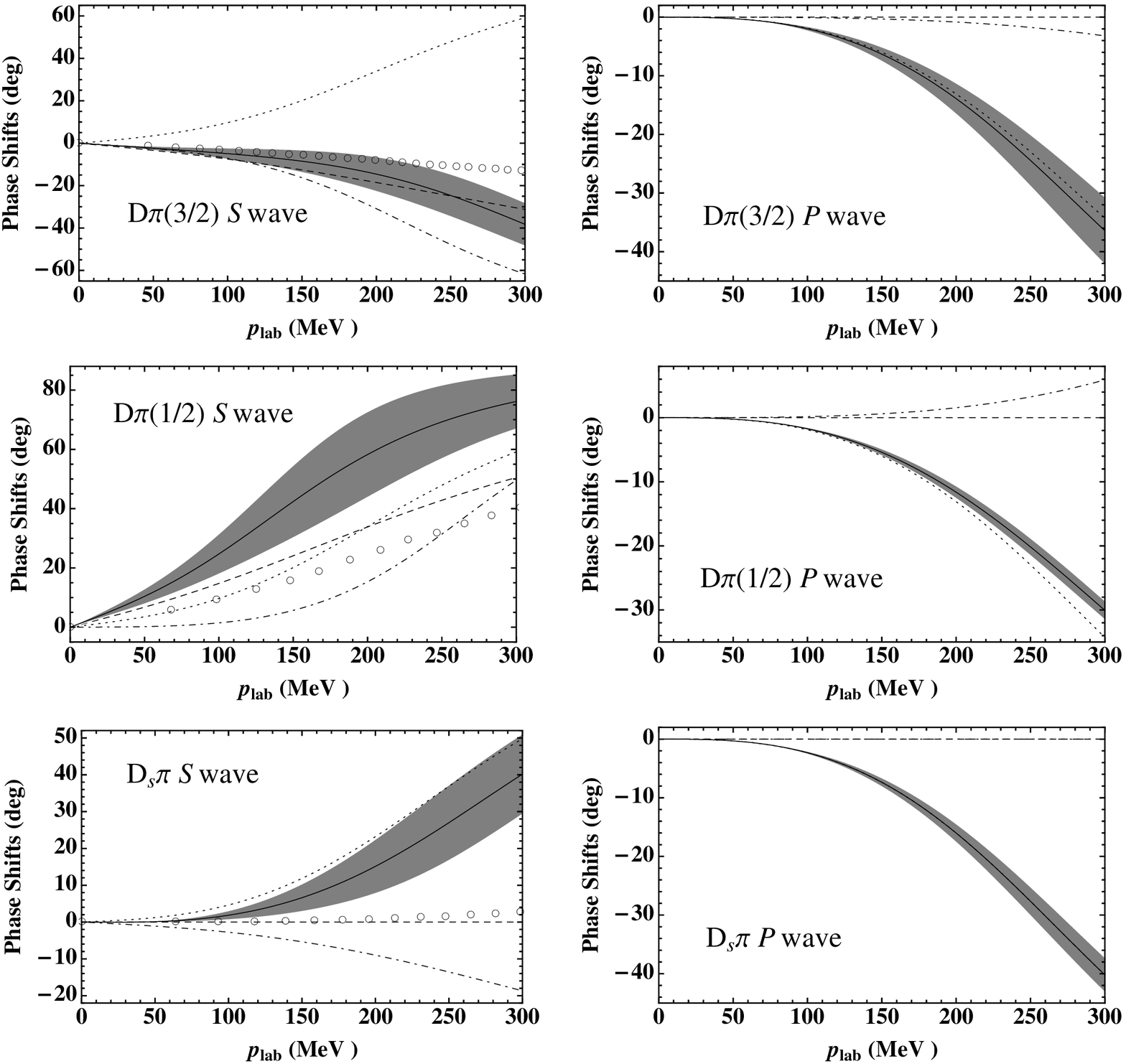}
\caption{\label{fig:piDphase}Predictions for the pion-$D$ meson
phase shifts versus the pion laboratory momentum at physical meson values.
The dashed, dotted, dash-dotted and solid lines denote the first-,
second-, third-order, and their total contributions, respectively. The open circles present the unitary results from refs.~\cite{guo2018,guo2022}.
The error bands are estimated from the statistical errors of the
LECs using the standard error propagation formula with the
correlations.}
\end{figure}

\begin{figure}[!t]
\centering
\includegraphics[height=12.5cm,width=12.5cm]{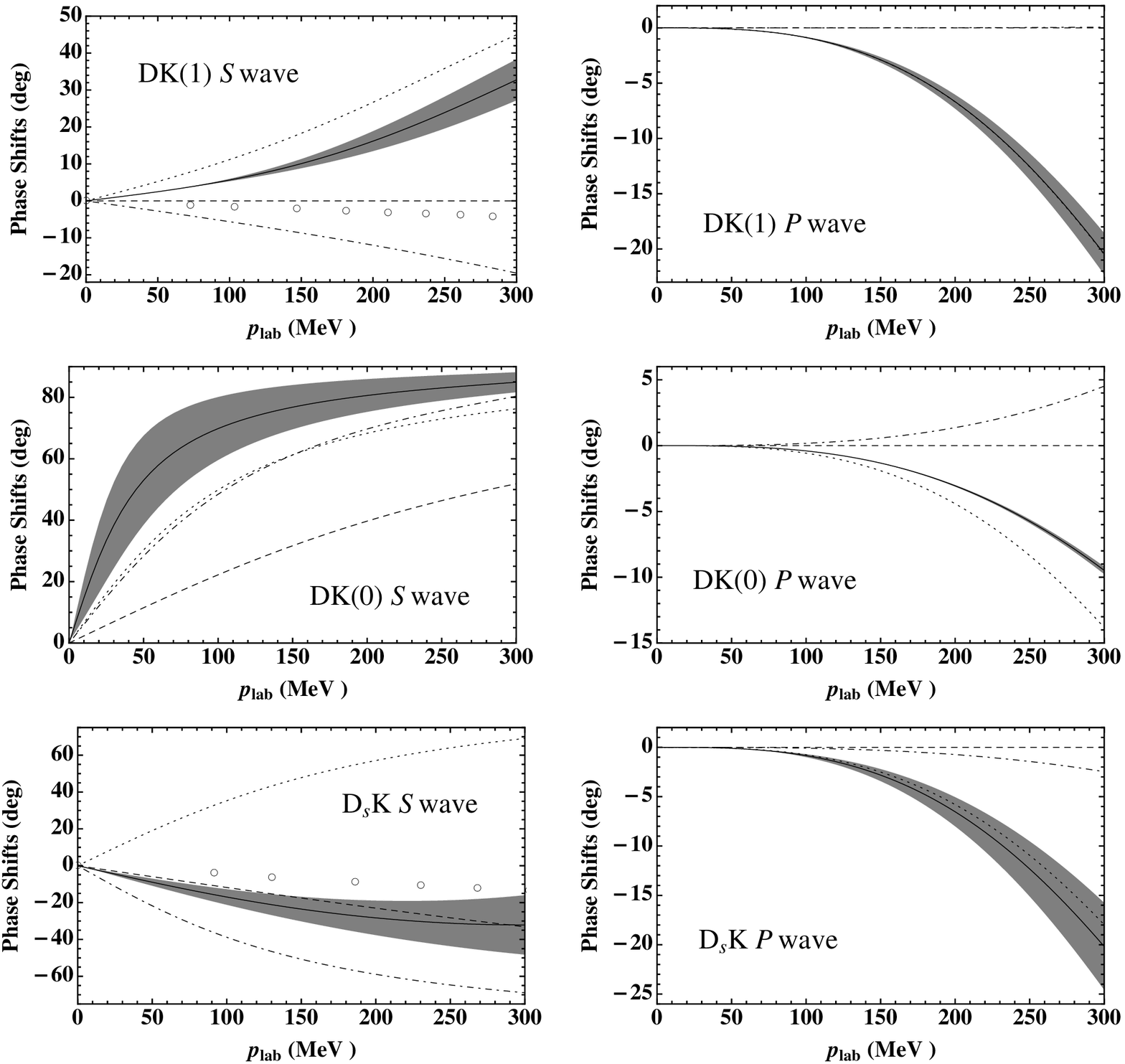}
\caption{\label{fig:KDphase}Predictions for the kaon-$D$ meson phase
shifts versus the kaon laboratory momentum at physical meson values. The
notation is the same as in Fig.~\ref{fig:piDphase}.}
\end{figure}

\begin{figure}[!t]
\centering
\includegraphics[height=12.5cm,width=12.5cm]{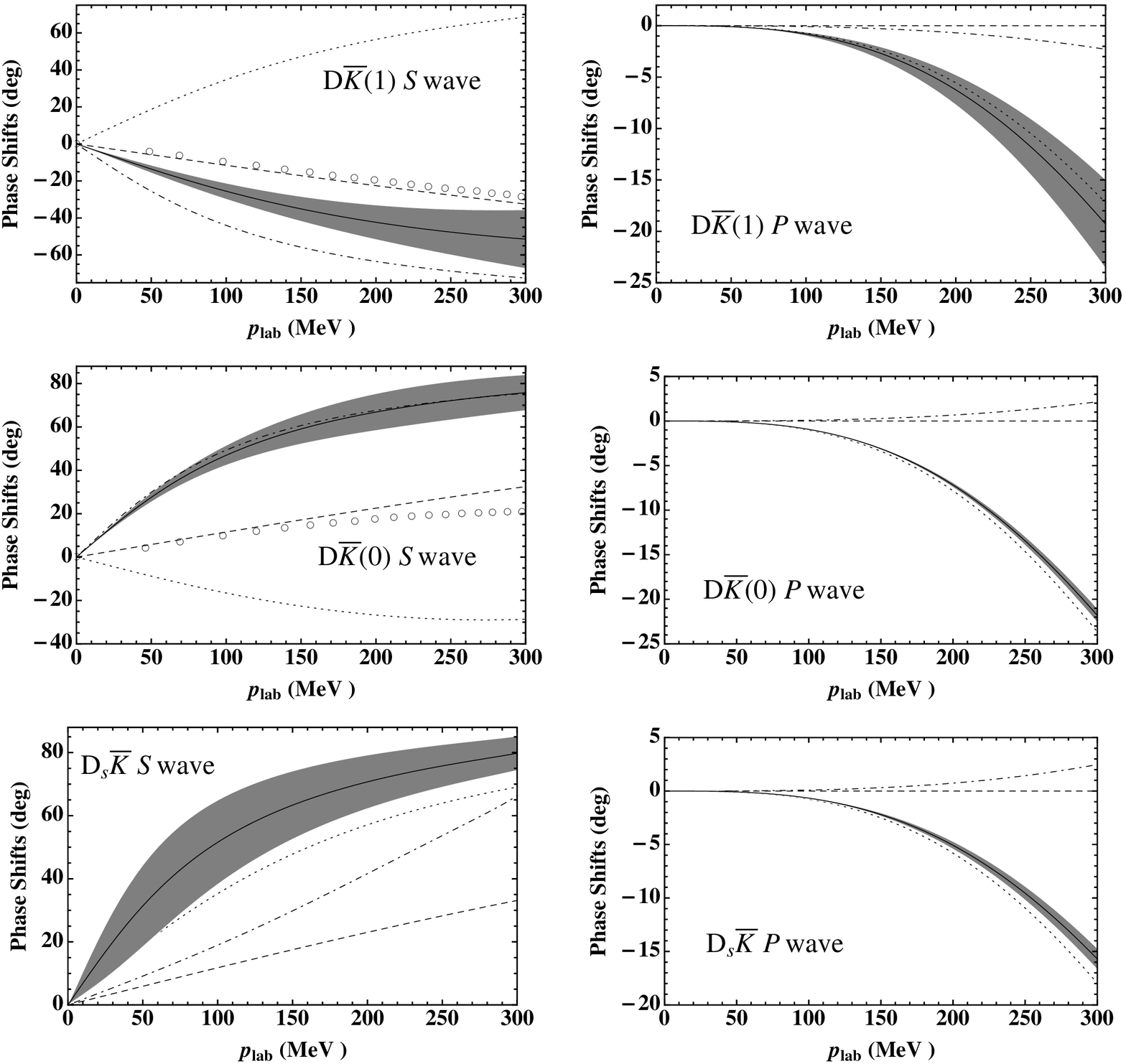}
\caption{\label{fig:KbarDphase}Predictions for the antikaon-$D$
meson phase shifts versus the antikaon laboratory momentum at
physical meson values. The notation is the same as in
Fig.~\ref{fig:piDphase}.}
\end{figure}

\begin{figure}[!t]
\centering
\includegraphics[height=8.4cm,width=12.5cm]{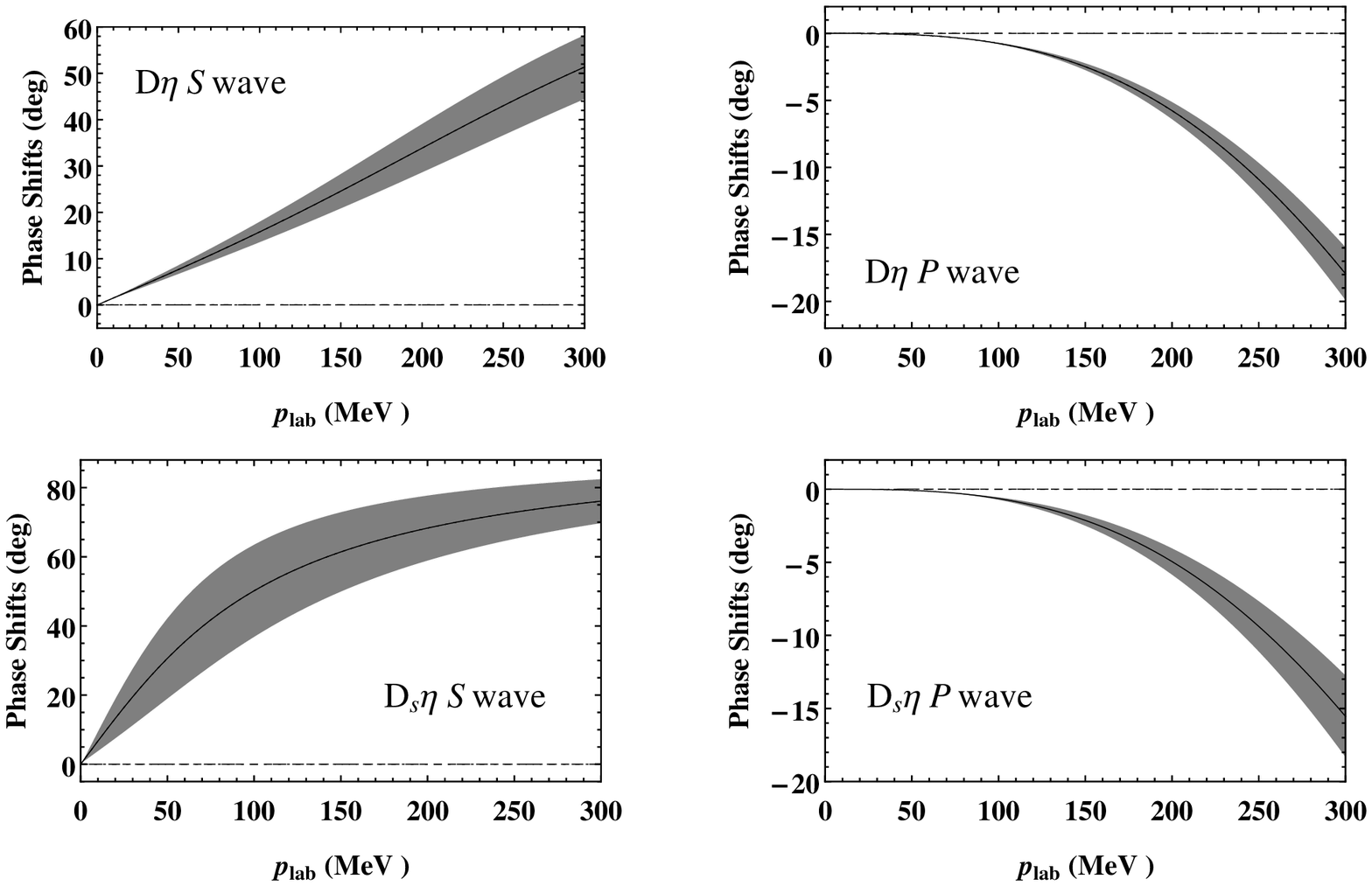}
\caption{\label{fig:etaDphase}Predictions for the eta-$D$ meson
phase shifts versus the eta laboratory momentum. The notation is the
same as in Fig.~\ref{fig:piDphase}.}
\end{figure}

In the following, we make predictions of the $S$- and $P$-wave phase
shifts for the eleven channels at the physical meson values using
the LECs from Table~\ref{fittingresult}. We use the values of the
physical parameters: $m_\pi=139.57\,\text{MeV}$,
$m_K=493.68\,\text{MeV}$, $f_\pi=92.07\,\text{MeV}$,
$f_K=110.03\,\text{MeV}$, $M_D=1869.66\,\text{MeV}$,
$M_{Ds}=1968.35\,\text{MeV}$ from PDG \cite{pdg2020}. The numerical
results of the phase shifts of the pion-, kaon-, antikaon-, and
eta-D meson scatterings are shown in Figs.~\ref{fig:piDphase},
\ref{fig:KDphase}, \ref{fig:KbarDphase}, and \ref{fig:etaDphase},
respectively. The error bands of the phase shifts in the total
contributions are estimated from the statistical errors of the LECs
using the standard error propagation formula with the correlations.
We can see that the bands from the LECs are not too large to be
unacceptable. The bands in the different orders are not given
because we do not determine the LECs at the corresponding orders,
although we present the values of the phase shifts from the
different orders. The convergence is not good for most of the
$S$-wave phase shifts, which is not surprising because it is
difficult to achieve good convergence at the third chiral order, as
in the case of the pion-nucleon scattering in ref.~\cite{fett1998}.
However, the $P$-wave phase shifts at the third chiral order are
much smaller than those at the second chiral order, which indicates
good convergence. In Figs.~\ref{fig:piDphase}, \ref{fig:KDphase},
\ref{fig:KbarDphase}, and \ref{fig:etaDphase}, we also show the
$S$-wave phase shifts calculated by the unitary method from
refs.~\cite{guo2018,guo2022} for making a more detailed comparison.

For the pion-$D$ meson phase shifts, we obtain the repulsions in the
$D\pi(I=3/2)$ $S$ wave and all $P$ waves, and attractions in the
$D\pi(I=1/2)$ and $D_s \pi$ $S$ waves. It is clear that there exist
no bound states or resonances in the channels with repulsions. The
attraction is weak below 200 MeV in the $D_s \pi$ $S$ wave. The attraction should not be strong enough to generate a bound state or resonance in this wave. The $D\pi(I=1/2)$ $S$ wave is particularly
interesting. The attraction exists at each order, and the total
attraction is very strong even below 200 MeV. The results from
lattice QCD simulations at nonphysical meson values support that there exists a bound state or resonance in this channel \cite{moir2016,greg2021}. However, the production of a bound state or resonance requires nonperturbative dynamics through an iterated method. We can see that the same direction is obtained between our calculations and the unitary results for the $S$-wave phase shifts. More detailed calculations including nonperturbative dynamics will be presented in forthcoming work.

For the kaon-D meson phase shifts, there are repulsions in the
$D_sK$ $S$ wave and all $P$ waves, and thus the bound state or
resonance cannot be dynamically generated in these waves. The
$DK(I=1)$ $S$ wave has weak attractions that cannot generate a
bound state or resonance. As expected, we obtain a strong attraction
in the $DK(I=0)$ $S$ wave, while the result from the unitary method has the opposite sign. This wave corresponds to the well-known
bound state $D_{s0}^{*}(2317)$. However, this exotic state has not been
directly obtained in our perturbative calculation. Nevertheless, it
is not difficult to obtain $D_{s0}^{*}(2317)$ by using an
iterated method (e.g., Schr\"{o}dinger equation) with the strong
attractive $DK$ interaction potential. The iterated method can
generate the bound state because the nonperturbative dynamics are
considered, as done in refs.~\cite{guo2018,guo2022}. A more detailed description of $D_{s0}^{*}(2317)$ will also be given in forthcoming work.

For the antikaon-D meson phase shifts, the $D\bar{K}(I=1)$ $S$ wave
and all $P$ waves have repulsions. Apparently, the bound state or
resonance cannot be found in these waves. Surprisingly, we obtain
strong attractions in both $D\bar{K}(I=0)$ and $D_s\bar{K}$ $S$
waves. The first-order contribution almost cancels the second-order contribution in the $D\bar{K}(I=0)$ $S$ wave, and the third-order
contribution dominates this wave. However, the total contribution is
still very large. The resulting strong attraction in this wave is
consistent with the lattice QCD result, which indicates that there
exists a virtual bound state in the $D\bar{K}(I=0)$ $S$ wave
\cite{cheu2021}. In the $D_s\bar{K}$ $S$ wave, the attraction is
obtained from each order. The total attraction is very strong and
supports the existence of a bound state. This wave corresponds to
the possible $D_{0}^{*}(2400)$ signal based on the coupled-channel
analysis of the $D\pi$, $D\eta$ and $D_s\bar{K}$ scattering
amplitudes in ref.~\cite{alba2017}. This is also consistent with the
strong attraction in the $D\pi(I=1/2)$ $S$ wave. There are also different signs between our perturbative calculation and the unitary result in this wave.

For the eta-D meson phase shifts, there are repulsions in all $P$
waves and strong attractions in all $S$ waves. The first- and
third-order contributions are almost zero in both $D\eta$ and
$D_s\eta$ $S$ waves since the tree amplitudes at the first- and
third-order are zero, and the one-loop amplitudes at the third order
are small. The $D\eta$ $S$ wave also corresponds to the
$D_{0}^{*}(2400)$ in the coupled-channel $D\pi$, $D\eta$ and
$D_s\bar{K}$ scattering amplitudes \cite{alba2017}. It is interesting
that the $D_s \eta$ $S$ wave also has strong attractions and
supports the existence of a bound state or resonance, while the opposite signs exist from the unity results of the refs.~\cite{guo2018,guo2022}. This will be further studied in future work.

From the phase shifts for the light pseudoscalar meson and heavy
meson scattering, we can see that there are repulsions in all $P$
waves, and the bound states or resonances cannot be dynamically
generated in these waves. However, we find that the phase shifts in
the $D\pi(I=1/2)$, $DK(I=0)$, $D\bar{K}(I=0)$, $D_s\bar{K}$, $D\eta$
and $D_s\eta$ $S$ waves are so strong that the bound states or
resonances may be generated dynamically in these channels.

\subsection{Scattering lengths and scattering volumes}

Finally, we calculate the scattering lengths for the $S$ waves and
the scattering volumes for the $P$ waves with Eq.~(\ref{thresholdpar}) at the physical meson values. Analytical expressions for the threshold parameters can be found in Appendix~\ref{thresholdpara}. The scattering
lengths are shown in Table~\ref{scenlen}, and the scattering volumes
are shown in Table~\ref{scenvolum}. The errors of the scattering lengths and the scattering volumes in our calculations are estimated from the statistical errors of the LECs using the error propagation formula with the correlations. Similarly, the errors at the different orders are not
given, although we present the values of the scattering lengths and
the scattering volumes from the different orders. Good convergence
is not achieved for the scattering lengths, while good convergence
is obtained for the scattering volumes.

The scattering lengths in the channels $D\pi(I=1/2)$, $DK(I=0)$,
$D\bar{K}(I=0)$, $D_s\bar{K}$, $D\eta$ and $D_s\eta$ have large
values. A bound state or resonance may be generated in these
channels. The other scattering lengths are either small or negative,
where a bound state or resonance cannot be dynamically generated. We
obtain a large positive value for the channel $DK(I=0)$, which
corresponds to $D_{s0}^{*}(2317)$. A channel with a bound state
should have a large negative scattering length, as obtained from
refs.~\cite{liu2013,guo2019}. The correct scattering length
for this channel was obtained in our previous work \cite{huan2022}
through the iterated method.

We obtain the negative values for the scattering volumes in all
channels. Therefore, a bound state or resonance cannot be generated
in the $P$ waves. The values from the first-order contributions are
zero, and the values from the third-order contributions are small.
Then, the second-order contributions dominate the total values. Good convergence is obtained for the scattering volumes at the third
chiral order.

\begin{table}[!t]
\centering
\resizebox{\textwidth}{!}{%
\begin{threeparttable}
\caption{\label{scenlen}Predictions of the scattering lengths for
the light pseudoscalar meson and $D$ meson at the physical meson values.
The scattering lengths are in units of fm.}
\begin{tabular}{ccccccccccccccccccc}
\midrule \toprule
Sca. Len. & $\mathcal{O}(p)$ & $\mathcal{O}(p^2)$ & $\mathcal{O}(p^3)$ &  \text{Total} & Liu2013 \cite{liu2013} & Guo2019 \cite{guo2019} & \\
\midrule
 $a_{(0,D\pi)}^{(3/2)}$ &$-0.24$& $0.23$ &$-0.16$ & $-0.17(6)$ & $-0.100(2)$  &  $-0.103^{+0.003}_{-0.003}$ & \\
\midrule
 $a_{(0,D\pi)}^{(1/2)}$ &$0.48$& $0.23$ & $0.00$  & $0.71(16)$ & $0.37^{+0.03}_{-0.02}$  &  $0.40^{+0.03}_{-0.02}$ &\\
\midrule
 $a_{(0,D_s \pi)}$ &$0.00$& $0.06$  & $-0.08$ &$-0.02(0)$& $-0.002(1)$ & $0.012^{+0.003}_{-0.003}$ &\\
\midrule
 $a_{(0,DK)}^{(1)}$  &$0.00$& $0.45$   & $-0.24$  & $0.21(1)$ & $0.07^{+0.03}_{-0.03}+i0.17^{+0.02}_{-0.01}$ & $-0.01^{+0.05}_{-0.03}+i0.39^{+0.04}_{-0.04}$ &\\
\midrule
 $a_{(0,DK)}^{(0)}$ &$1.01$& $2.90$   & $2.65$ & $6.57(318)$ & $-0.84^{+0.17}_{-0.22}$ & $-1.51^{+0.72}_{-2.35}$ & \\
\midrule
$a_{(0,D_s K)}$ & $-0.51$ & $1.69$   & $-1.96$ & $-0.78(19)$ & $-0.18(1)$ & $-0.20^{+0.01}_{-0.01}$ & \\
\midrule
 $a_{(0,D\bar{K})}^{(1)}$ & $-0.51$ & $1.67$  & $-2.38$ & $-1.21(19)$ & $-0.20(1)$ & $-0.20^{+0.01}_{-0.01}$ & \\
\midrule
$a_{(0,D\bar{K})}^{(0)}$  & $0.51$ & $-0.78$   & $2.86$  &  $2.58(19)$ & $0.84(15)$ & $21.9$ & \\
\midrule
 $a_{(0,D_s \bar{K})}$ & $0.51$ &  $1.69$  & $0.78$ & $2.98(161)$ & $-0.09^{+0.06}_{-0.05}+i0.44^{+0.05}_{-0.05}$ & $-0.57^{+0.06}_{-0.04}+i0.35^{+0.08}_{-0.07}$ & \\
\midrule
$a_{(0,D\eta)}$ & $0.00$ & $0.67$  & $0.00$ & $0.67(9)$ &  & $0.29^{+0.15}_{-0.22}+i0.61^{+0.30}_{-0.26}$  & \\
\midrule
$a_{(0,D_s \eta)}$ & $0.00$ & $2.97$  & $0.00$ & $2.97(139)$ &  & $-0.39^{+0.05}_{-0.03}+i0.06^{+0.02}_{-0.02}$ &  \\
\bottomrule \midrule
\end{tabular}
\end{threeparttable}}%
\end{table}

\begin{table}[!t]
\centering
\begin{threeparttable}
\caption{\label{scenvolum}Predictions of the scattering volumes for
the light pseudoscalar meson and $D$ meson at the physical meson values.
The scattering volumes are in units of $\text{fm}^3$. Note that, the
values for the $\mathcal{O}(p)$ in all channels are zero, and are
not shown.}
\begin{tabular}{ccccccccccccccccccc}
\midrule \toprule
Sca. Vol. & $\mathcal{O}(p^2)$ & $\mathcal{O}(p^3)$ &  \text{Total} & \\
\midrule
 $a_{(1,D\pi)}^{(3/2)}$ & $-0.33$ &$-0.01$ & $-0.34(6)$ & \\
\midrule
 $a_{(1,D\pi)}^{(1/2)}$ & $-0.33$ & $0.02$  & $-0.31(4)$ & \\
\midrule
 $a_{(1,D_s \pi)}$ & $-0.40$  & $0.00$ &$-0.40(4)$&\\
\midrule
 $a_{(1,DK)}^{(1)}$  &$-0.24$   & $0.00$  & $-0.24(2)$ &\\
\midrule
 $a_{(1,DK)}^{(0)}$ & $-0.16$   & $0.05$ & $-0.11(0)$ & \\
\midrule
$a_{(1,D_s K)}$ &  $-0.20$   & $-0.02$ & $-0.22(5)$ &  \\
\midrule
 $a_{(1,D\bar{K})}^{(1)}$ &  $-0.20$  & $-0.02$ & $-0.22(5)$ &  \\
\midrule
$a_{(1,D\bar{K})}^{(0)}$  &  $-0.28$   & $0.02$  &  $-0.26(1)$ &  \\
\midrule
 $a_{(1,D_s \bar{K})}$ &   $-0.20$  & $0.02$ & $-0.18(1)$ &  \\
\midrule
$a_{(1,D\eta)}$ &  $-0.22$  & $0.00$ & $-0.22(3)$ & \\
\midrule
$a_{(1,D_s \eta)}$ & $-0.18$  & $0.00$ & $-0.18(3)$ &  \\
\bottomrule \midrule
\end{tabular}
\end{threeparttable}
\end{table}

\section{Summary}

In summary, we have calculated the complete $T$ matrices of the
elastic light pseudoscalar meson and heavy meson scattering up to
the third order in HMChPT. We fitted the phase shifts and the
scattering lengths from lattice QCD at nonphysical meson values to
determine the LECs. This led to a good description of the phase
shifts below the 200 MeV pion/kaon momentum and the scattering
lengths at the nonphysical meson values for the channels excluding a
bound state or resonance. We also obtained the LEC uncertainties and
their mutual correlations through statistical regression analysis.
We predicted the $S$- and $P$-wave phase shifts for light
pseudoscalar meson and heavy meson scattering using these LECs at
the physical meson values. We found that the phase shifts in the
$D\pi(I=1/2)$, $DK(I=0)$, $D\bar{K}(I=0)$, $D_s\bar{K}$, $D\eta$ and
$D_s\eta$ $S$ waves are strong enough to generate a bound state or
resonance. The channel $DK(I=0)$ corresponds to the well-known
$D_{s0}^{*}(2317)$. The coupled channels $D\pi(I=1/2)$, $D_s\bar{K}$
and $D\eta$ may correspond to $D_{0}^{*}(2400)$. The channels
$D\bar{K}(I=0)$ and $D_s\eta$ may generate the respective bound
state or resonance. However, as expected, we cannot obtain directly
a bound state or resonance in our perturbative calculations. This
issue can be successfully solved by the nonperturbative method, and
the calculations including the nonperturbative dynamics will be
presented in a forthcoming work. The $P$ wave phase shifts in all
channels are repulsive, and the bound states or resonances cannot be
dynamically generated in these waves. We also predicted the
scattering lengths and the scattering volumes using the LECs at the
physical meson values. The scattering lengths also have large values
in the channels $D\pi(I=1/2)$, $DK(I=0)$, $D\bar{K}(I=0)$,
$D_s\bar{K}$, $D\eta$ and $D_s\eta$, which indicate that a bound
state or resonance may be generated in these channels. However, the
correct scattering lengths for these channels should be obtained
through the iterated method. We obtained negative values for the
scattering volumes in all channels, and a bound state or resonance
cannot be generated in the $P$ waves. In addition, we obtained good
convergence for the scattering volumes at the third chiral order. In
order to study the bound states or resonances directly, the
calculation including the nonperturbative dynamics is necessary. We
hope our present calculations contribute to the investigations on
the heavy meson-heavy meson interactions in HMChPT.

\section*{Acknowledgments}
This work is supported by the National Natural Science Foundation of
China under Grants No. 11975033, No. 12070131001 and No. 12147127,
and China Postdoctoral Science Foundation (Grant No. 2021M700251). We thank Xiao-Yu Guo (Beijing University of Technology), Feng-Kun Guo (Beijing, Institute of Theoretical Physics), Jing Ou-Yang (Yunnan University) and Norbert Kaiser (Technische Universit\"{a}t München) for very helpful discussions.

\appendix\markboth{Appendix}{Appendix}
\renewcommand{\thesection}{\Alph{section}}
\numberwithin{equation}{section} \numberwithin{table}{section}

\section{$D_{s0}^{*}(2317)$ contribution}
\label{D2317contr}
\begin{figure}[h]
\centering
\includegraphics[height=2cm,width=5cm]{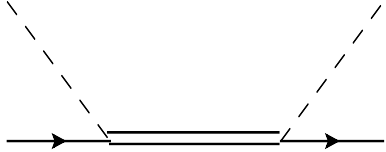}
\caption{\label{fig:feynmanD2317}The leading $D_{s0}^{*}(2317)$-exchange Born-term contribution. The double solid, solid and dashed lines represent $D_{s0}^{*}(2317)$, $D$ meson, and kaon, respectively. The crossed diagram is not shown.}
\end{figure}

Denoting the $D_{s0}^{*}(2317)$ by $D_R$, the leading-order effective Lagrangian with $D_R$ as explicit degree of freedom reads
\begin{align}
\label{lagD2317}
 \mathcal{L}_{D_R}=\bar{D}_R(iv\cdot\partial-M_{R}+M_D)D_R+(g_{R}\left\langle \bar{D}_R v\cdot A\,D \right\rangle+\text{h.c.})
\end{align}
where $A=\frac{\partial_\mu K}{f}$ with $K=(K^+,K^0)^{\text{T}}\,\text{or}\,(K^{-},\bar{K}^{0})^{\text{T}}$, and $D=(D^0,D^+)$. The leading $D_{s0}^{*}(2317)$-exchange Born-term contribution resulting from Fig.~\ref{fig:feynmanD2317} reads
\begin{align}
\label{tmaD2317}
T_{KD}^{(0)}=\frac{2w_K^2g_R^2}{f_K^2}\Big(\frac{1}{M_R-M_D-w_K}+\frac{1}{M_R-M_D+w_K}\Big).
\end{align}
Putting the LO, NLO, N2LO amplitudes and the Eq.~(\ref{tmaD2317}) together in $KD(I=0)$ channel, we can obtain the complete $s$-wave scattering lengths
\begin{align}
\label{aD2317}
a_{KD}^{(0)}=&\frac{M_D}{8\pi(M_D+m_K)f_K^2}\Big[2m_K+2(4c_0+4c_1+c_2+c_3+c_4+c_5)m_K^2+4(4\bar{\kappa}_1+\bar{\kappa}_2+\bar{\kappa}_3)m_K^3\nonumber\\
&+\frac{3m_K^2}{8\pi^2f_K^2}\Big(m_K-\sqrt{m_\eta^2-m_K^2}\arccos\frac{-m_K}{m_\eta}\Big)\nonumber\\
&+2m_K^2g_R^2\Big(\frac{1}{M_R-M_D-m_K}+\frac{1}{M_R-M_D+m_K}\Big)\Big].
\end{align}

\section{Threshold parameters}
\label{thresholdpara}
In this appendix, we give the analytical expressions for the threshold parameters up to third order. These read:
\begin{align}
\label{a0piD1c2}
a_{(0,\pi D)}^{(1/2)}=&\frac{M_D}{8\pi(M_D+m_\pi)f_\pi^2}\Big[2m_\pi+(8c_0+4c_1+2c_2+c_3+2c_4+c_5)m_\pi^2+4(4\bar{\kappa}_1+\bar{\kappa}_2+\bar{\kappa}_3)m_\pi^3\nonumber\\
&-\frac{m_\pi^2}{16\pi^2f_\pi^2}\Big(6m_\pi+\sqrt{m_K^2-m_\pi^2}\arccos\frac{m_\pi}{m_K}-3\sqrt{m_K^2-m_\pi^2}\arccos\frac{-m_\pi}{m_K}\Big)\Big],
\end{align}
\begin{align}
\label{a1piD1c2}
a_{(1,\pi D)}^{(1/2)}=\frac{M_D}{16\pi(M_D+m_\pi)f_\pi^2}\Big(-\frac{4}{3}c_4-\frac{2}{3}c_5-\frac{8}{3}\bar{\kappa}_3m_\pi+\frac{7}{864\pi^2f_\pi^2}m_\pi\Big),
\end{align}
\begin{align}
\label{a0piD3c2}
a_{(0,\pi D)}^{(3/2)}=&\frac{M_D}{8\pi(M_D+m_\pi)f_\pi^2}\Big[-m_\pi+(8c_0+4c_1+2c_2+c_3+2c_4+c_5)m_\pi^2-2(4\bar{\kappa}_1+\bar{\kappa}_2+\bar{\kappa}_3)m_\pi^3\nonumber\\
&-\frac{m_\pi^2}{16\pi^2f_\pi^2}\Big(3m_\pi+2\sqrt{m_K^2-m_\pi^2}\arccos\frac{m_\pi}{m_K}\Big)\Big],
\end{align}
\begin{align}
\label{a1piD3c2}
a_{(1,\pi D)}^{(3/2)}=\frac{M_D}{16\pi(M_D+m_\pi)f_\pi^2}\Big(-\frac{4}{3}c_4-\frac{2}{3}c_5+\frac{4}{3}\bar{\kappa}_3m_\pi-\frac{7}{432\pi^2f_\pi^2}m_\pi\Big),
\end{align}
\begin{align}
\label{a0piDs}
a_{(0,\pi D_s)}=&\frac{M_{D_s}}{8\pi(M_{D_s}+m_\pi)f_\pi^2}\Big[2(4c_0+c_2+c_4)m_\pi^2-\frac{m_\pi^2}{8\pi f_\pi^2}\sqrt{m_K^2-m_\pi^2}\,\Big],
\end{align}
\begin{align}
\label{a1piDs}
a_{(1,\pi D_s)}=\frac{M_{D_s}}{16\pi(M_{D_s}+m_\pi)f_\pi^2}\Big(-\frac{4}{3}c_4\Big),
\end{align}
\begin{align}
\label{a0KD1}
a_{(0,K D)}^{(1)}=&\frac{M_D}{8\pi(M_D+m_K)f_K^2}\Big\{2(4c_0+c_2+c_4)m_K^2\nonumber\\
&+\frac{m_K^2}{8\pi^2f_K^2}\Big[\sqrt{m_K^2-m_\pi^2}\Big(i\pi-\ln\frac{m_K+\sqrt{m_K^2-m_\pi^2}}{m_\pi}\Big) \Big]\Big\},
\end{align}
\begin{align}
\label{a1KD1}
a_{(1,K D)}^{(1)}=\frac{M_{D}}{16\pi(M_{D}+m_K)f_K^2}\Big(-\frac{4}{3}c_4+\frac{m_K}{288\pi^2f_K^2}\Big),
\end{align}
\begin{align}
\label{a0KD0}
a_{(0,K D)}^{(0)}=&\frac{M_D}{8\pi(M_D+m_K)f_K^2}\Big[2m_K+2(4c_0+4c_1+c_2+c_3+c_4+c_5)m_K^2+4(4\bar{\kappa}_1+\bar{\kappa}_2+\bar{\kappa}_3)m_K^3\nonumber\\
&+\frac{3m_K^2}{8\pi^2f_K^2}\Big(m_K-\sqrt{m_\eta^2-m_K^2}\arccos\frac{-m_K}{m_\eta}\Big)\Big],
\end{align}
\begin{align}
\label{a1KD1}
a_{(1,K D)}^{(0)}=\frac{M_{D}}{16\pi(M_{D}+m_K)f_K^2}\Big(-\frac{4}{3}c_4-\frac{4}{3}c_5-\frac{8\bar{\kappa}_3}{3}m_K-\frac{23}{864\pi^2f_K^2}m_K\Big),
\end{align}
\begin{align}
\label{a0KDs}
a_{(0,K D_s)}=&\frac{M_{D_s}}{8\pi(M_{D_s}+m_K)f_K^2}\Big[-m_K+(8c_0+4c_1+2c_2+c_3+2c_4+c_5)m_K^2-(8\bar{\kappa}_1+2\bar{\kappa}_2\nonumber\\
&+2\bar{\kappa}_3)m_K^3-\frac{3m_K^2}{16\pi^2f_K^2}\Big(m_K+\sqrt{m_\eta^2-m_K^2}\arccos\frac{m_K}{m_\eta}\nonumber\\
&-\sqrt{m_K^2-m_\pi^2}\ln\frac{m_K+\sqrt{m_K^2-m_\pi^2}}{m_\pi}\Big)\Big],
\end{align}
\begin{align}
\label{a1KDs}
a_{(1,K D_s)}=\frac{M_{D_s}}{16\pi(M_{D_s}+m_K)f_K^2}\Big(-\frac{4}{3}c_4-\frac{2}{3}c_5+\frac{4\bar{\kappa}_3}{3}m_K+\frac{7}{864\pi^2f_K^2}m_K\Big),
\end{align}
\begin{align}
\label{a0KbarD1}
a_{(0,\bar{K} D)}^{(1)}=&\frac{M_D}{8\pi(M_D+m_K)f_K^2}\Big[-m_K+(8c_0+4c_1+2c_2+c_3+2c_4+c_5)m_K^2-(8\bar{\kappa}_1+2\bar{\kappa}_2\nonumber\\
&+2\bar{\kappa}_3)m_K^3+\frac{m_K^2}{16\pi^2f_K^2}\Big(-3m_K-3\sqrt{m_\eta^2-m_K^2}\arccos\frac{m_K}{m_\eta}\nonumber\\
&+\sqrt{m_K^2-m_\pi^2}\ln\frac{m_K+\sqrt{m_K^2-m_\pi^2}}{m_\pi} \Big)\Big],
\end{align}
\begin{align}
\label{a1KbarD1}
a_{(1,\bar{K} D)}^{(1)}=\frac{M_{D}}{16\pi(M_{D}+m_K)f_K^2}\Big(-\frac{4}{3}c_4-\frac{2}{3}c_5+\frac{4\bar{\kappa}_3}{3}m_K+\frac{5}{432\pi^2f_K^2}m_K\Big),
\end{align}
\begin{align}
\label{a0KbarD0}
a_{(0,\bar{K} D)}^{(0)}=&\frac{M_D}{8\pi(M_D+m_K)f_K^2}\Big[m_K+(8c_0-4c_1+2c_2-c_3+2c_4-c_5)m_K^2+(8\bar{\kappa}_1+2\bar{\kappa}_2\nonumber\\
&+2\bar{\kappa}_3)m_K^3+\frac{3m_K^2}{16\pi^2f_K^2}\Big(m_K+\sqrt{m_\eta^2-m_K^2}\arccos\frac{m_K}{m_\eta}\nonumber\\
&+\sqrt{m_K^2-m_\pi^2}\ln\frac{m_K+\sqrt{m_K^2-m_\pi^2}}{m_\pi} \Big)\Big],
\end{align}
\begin{align}
\label{a1KbarD0}
a_{(1,\bar{K} D)}^{(0)}=\frac{M_{D}}{16\pi(M_{D}+m_K)f_K^2}\Big(-\frac{4}{3}c_4+\frac{2}{3}c_5-\frac{4\bar{\kappa}_3}{3}m_K-\frac{1}{54\pi^2f_K^2}m_K\Big),
\end{align}
\begin{align}
\label{a0KbarDs}
a_{(0,\bar{K} D_s)}=&\frac{M_{D_s}}{8\pi(M_{D_s}+m_K)f_K^2}\Big\{m_K+(8c_0+4c_1+2c_2+c_3+2c_4+c_5)m_K^2+(8\bar{\kappa}_1+2\bar{\kappa}_2\nonumber\\
&+2\bar{\kappa}_3)m_K^3-\frac{3m_K^2}{16\pi^2f_K^2}\Big[m_K-\sqrt{m_\eta^2-m_K^2}\arccos\frac{-m_K}{m_\eta}\nonumber\\
&+\Big(i\pi-\sqrt{m_K^2-m_\pi^2}\ln\frac{m_K+\sqrt{m_K^2-m_\pi^2}}{m_\pi}\Big) \Big]\Big\},
\end{align}
\begin{align}
\label{a1KbarDs}
a_{(1,\bar{K} D_s)}=\frac{M_{D_s}}{16\pi(M_{D_s}+m_K)f_K^2}\Big(-\frac{4}{3}c_4-\frac{2}{3}c_5-\frac{4\bar{\kappa}_3}{3}m_K-\frac{7}{864\pi^2f_K^2}m_K\Big),
\end{align}
\begin{align}
\label{a0etaD}
a_{(0,\eta D)}=&\frac{M_{D}}{8\pi(M_{D}+m_\eta)f_\eta^2}\Big\{\frac{1}{9}[4(24c_0+6c_2+c_3+6c_4+c_5)m_K^2-(24c_0-12c_1+6c_2+c_3\nonumber\\
&+6c_4+c_5)m_\pi^2]+\frac{3m_\eta^2}{16\pi^2f_\eta^2}(i\pi)\Big\},
\end{align}
\begin{align}
\label{a1etaD}
a_{(1,\eta D)}=\frac{M_{D}}{16\pi(M_{D}+m_\eta)f_\eta^2}\Big(-\frac{4}{3}c_4-\frac{2}{9}c_5\Big),
\end{align}
\begin{align}
\label{a0etaDs}
a_{(0,\eta D_s)}=&\frac{M_{D_s}}{8\pi(M_{D_s}+m_\eta)f_\eta^2}\Big\{\frac{1}{9}[8(12c_0+12c_1+3c_2+2c_3+3c_4+2c_5)m_K^2-2(12c_0+24c_1\nonumber\\
&+3c_2+2c_3+3c_4+2c_5)m_\pi^2]+\frac{3m_\eta^2}{8\pi^2f_\eta^2}(i\pi)\Big\},
\end{align}
\begin{align}
\label{a1etaDs}
a_{(1,\eta D_s)}=\frac{M_{D_s}}{16\pi(M_{D_s}+m_\eta)f_\eta^2}\Big(-\frac{4}{3}c_4-\frac{8}{9}c_5\Big).
\end{align}

\bibliographystyle{unsrt}
\bibliography{mesonheavymesonscattering}

\providecommand{\noopsort}[1]{}\providecommand{\singleletter}[1]{#1}%
\begin{thebibliography}{10}

\bibitem{wein1979}
S.~Weinberg.
\newblock {Phenomenological Lagrangians}.
\newblock {\em Physica A}, 96:327--340, 1979.
\newblock \url{https://doi.org/10.1016/0378-4371(79)90223-1}.

\bibitem{sche2012}
S.~Scherer and M.~R. Schindler.
\newblock {A primer for chiral perturbation theory}.
\newblock {\em Lect. Notes Phys.}, 830:1--338, 2012.
\newblock \url{https://doi.org/10.1007/978-3-642-19254-8}.

\bibitem{mach2011}
R.~Machleidt and D.~R. Entem.
\newblock {Chiral effective field theory and nuclear forces}.
\newblock {\em Phys. Rept.}, 503:1--75, 2011.
\newblock \url{https://doi.org/10.1016/j.physrep.2011.02.001}.

\bibitem{gass1988}
J.~Gasser, M.~E. Sainio, and A.~Svarc.
\newblock {Nucleons with Chiral Loops}.
\newblock {\em Nucl. Phys. B}, 307:779--853, 1988.
\newblock \url{https://doi.org/10.1016/0550-3213(88)90108-3}.

\bibitem{jenk1991}
E.~E. Jenkins and A.~V. Manohar.
\newblock {Baryon chiral perturbation theory using a heavy fermion Lagrangian}.
\newblock {\em Phys. Lett. B}, 255:558--562, 1991.
\newblock \url{https://doi.org/10.1016/0370-2693(91)90266-S}.

\bibitem{bern1992}
V.~Bernard, N.~Kaiser, J.~Kambor, and U.-G. Mei{\ss}ner.
\newblock {Chiral structure of the nucleon}.
\newblock {\em Nucl. Phys. B}, 388:315--345, 1992.
\newblock \url{https://doi.org/10.1016/0550-3213(92)90615-I}.

\bibitem{ordo1992}
C.~Ordóñez and U.~{van Kolck}.
\newblock Chiral lagrangians and nuclear forces.
\newblock {\em Phys. Lett. B}, 291:459--464, 1992.
\newblock \url{https://doi.org/10.1016/0370-2693(92)91404-W}.

\bibitem{epel1998}
E.~Epelbaoum, W.~Glöckle, and U.-G. Meißner.
\newblock Nuclear forces from chiral lagrangians using the method of unitary
  transformation ({I}): Formalism.
\newblock {\em Nucl. Phys. A}, 637:107--134, 1998.
\newblock \url{https://doi.org/10.1016/S0375-9474(98)00220-6}.

\bibitem{fett1998}
N.~Fettes, U.-G. Mei{\ss}ner, and S.~Steininger.
\newblock {Pion-nucleon scattering in chiral perturbation theory (I): Isospin
  symmetric case}.
\newblock {\em Nucl. Phys. A}, 640:199--234, 1998.
\newblock \url{https://doi.org/10.1016/S0375-9474(98)00452-7}.

\bibitem{fett2000}
N.~Fettes and U.-G. Mei{\ss}ner.
\newblock {Pion nucleon scattering in chiral perturbation theory (II): Fourth
  order calculation}.
\newblock {\em Nucl. Phys. A}, 676:311, 2000.
\newblock \url{https://doi.org/10.1016/S0375-9474(00)00199-8}.

\bibitem{kais19971}
Norbert Kaiser, R.~Brockmann, and W.~Weise.
\newblock {Peripheral nucleon-nucleon phase shifts and chiral symmetry}.
\newblock {\em Nucl. Phys. A}, 625:758--788, 1997.
\newblock \url{https://doi.org/10.1016/S0375-9474(97)00586-1}.

\bibitem{kang2014}
X.-W. Kang, J.~Haidenbauer, and U.-G. Mei\ss{}ner.
\newblock {Antinucleon-nucleon interaction in chiral effective field theory}.
\newblock {\em JHEP}, 02:113, 2014.
\newblock \url{https://doi.org/10.1007/JHEP02(2014)113}.

\bibitem{ente2015}
D.~R. Entem, N.~Kaiser, R.~Machleidt, and Y.~Nosyk.
\newblock {Peripheral nucleon-nucleon scattering at fifth order of chiral
  perturbation theory}.
\newblock {\em Phys. Rev. C}, 91:014002, 2015.
\newblock \url{https://doi.org/10.1103/PhysRevC.91.014002}.

\bibitem{kais2020}
N.~Kaiser.
\newblock {Density-dependent NN interaction from subsubleading chiral 3N
  forces: Intermediate-range contributions}.
\newblock {\em Phys. Rev. C}, 101:014001, 2020.
\newblock \url{https://doi.org/10.1103/PhysRevC.101.014001}.

\bibitem{kais2001}
N.~Kaiser.
\newblock {Chiral corrections to kaon nucleon scattering lengths}.
\newblock {\em Phys. Rev. C}, 64:045204, 2001.
\newblock \url{https://doi.org/10.1103/PhysRevC.64.045204}.

\bibitem{liu20071}
Y.-R. Liu and S.-L. Zhu.
\newblock {Meson-baryon scattering lengths in HB$\chi$PT}.
\newblock {\em Phys. Rev. D}, 75:034003, 2007.
\newblock \url{https://doi.org/10.1103/PhysRevD.75.034003}.

\bibitem{haid2013}
J.~Haidenbauer, S.~Petschauer, N.~Kaiser, U.-G. Meißner, A.~Nogga, and
  W.~Weise.
\newblock Hyperon–nucleon interaction at next-to-leading order in chiral
  effective field theory.
\newblock {\em Nucl. Phys. A}, 915:24--58, 2013.
\newblock \url{https://doi.org/10.1016/j.nuclphysa.2013.06.008}.

\bibitem{huan2015}
B.-L. Huang and Y.-D. Li.
\newblock {Kaon-nucleon scattering to one-loop order in heavy baryon chiral
  perturbation theory}.
\newblock {\em Phys. Rev. D}, 92:114033, 2015.
\newblock \url{https://doi.org/10.1103/PhysRevD.92.114033}.

\bibitem{huan2017}
B.-L. Huang, J.-S. Zhang, Y.-D. Li, and N.~Kaiser.
\newblock {Meson-baryon scattering to one-loop order in heavy baryon chiral
  perturbation theory}.
\newblock {\em Phys. Rev. D}, 96:016021, 2017.
\newblock \url{https://doi.org/10.1103/PhysRevD.96.016021}.

\bibitem{huan20201}
B.-L. Huang and J.~Ou-Yang.
\newblock {Pion-nucleon scattering to $\mathcal{O}(p^3)$ in heavy baryon SU(3)
  chiral perturbation theory}.
\newblock {\em Phys. Rev. D}, 101:056021, 2020.
\newblock \url{https://doi.org/10.1103/PhysRevD.101.056021}.

\bibitem{huan20202}
B.-L. Huang.
\newblock {Pion-nucleon scattering to order $p^4$ in SU(3) heavy baryon chiral
  perturbation theory}.
\newblock {\em Phys. Rev. D}, 102:116001, 2020.
\newblock \url{https://doi.org/10.1103/PhysRevD.102.116001}.

\bibitem{huan2021}
B.-L. Huang, J.-B. Cheng, and S.-L. Zhu.
\newblock {Peripheral nucleon-nucleon scattering at next-to-next-to-leading
  order in SU(3) heavy baryon chiral perturbation theory}.
\newblock {\em Phys. Rev. D}, 104:116030, 2021.
\newblock \url{https://doi.org/10.1103/PhysRevD.104.116030}.

\bibitem{bech1999}
T.~Becher and H.~Leutwyler.
\newblock {Baryon chiral perturbation theory in manifestly Lorentz invariant
  form}.
\newblock {\em Eur. Phys. J. C}, 9:643--671, 1999.
\newblock \url{https://doi.org/10.1007/PL00021673}.

\bibitem{gege1999}
J.~Gegelia and G.~Japaridze.
\newblock {Matching heavy particle approach to relativistic theory}.
\newblock {\em Phys. Rev. D}, 60:114038, 1999.
\newblock \url{https://doi.org/10.1103/PhysRevD.60.114038}.

\bibitem{fuch2003}
T.~Fuchs, J.~Gegelia, G.~Japaridze, and S.~Scherer.
\newblock {Renormalization of relativistic baryon chiral perturbation theory
  and power counting}.
\newblock {\em Phys. Rev. D}, 68:056005, 2003.
\newblock \url{https://doi.org/10.1103/PhysRevD.68.056005}.

\bibitem{schi2007}
M.~R. Schindler, T.~Fuchs, J.~Gegelia, and S.~Scherer.
\newblock {Axial, induced pseudoscalar, and pion-nucleon form-factors in
  manifestly Lorentz-invariant chiral perturbation theory}.
\newblock {\em Phys. Rev. C}, 75:025202, 2007.
\newblock \url{https://doi.org/10.1103/PhysRevC.75.025202}.

\bibitem{geng2008}
L.~S. Geng, J.~Martin~Camalich, L.~Alvarez-Ruso, and M.~J. Vicente~Vacas.
\newblock {Leading SU(3)-breaking corrections to the baryon magnetic moments in
  chiral perturbation theory}.
\newblock {\em Phys. Rev. Lett.}, 101:222002, 2008.
\newblock \url{https://doi.org/10.1103/PhysRevLett.101.222002}.

\bibitem{alar2012}
J.~M. Alarc\'{o}n, J.~Martin~Camalich, and J.~A. Oller.
\newblock {The chiral representation of the $\pi N$ scattering amplitude and
  the pion-nucleon sigma term}.
\newblock {\em Phys. Rev. D}, 85:051503, 2012.
\newblock \url{https://doi.org/10.1103/PhysRevD.85.051503}.

\bibitem{ren2012}
X.~L. Ren, L.~S. Geng, J.~Martin~Camalich, J.~Meng, and H.~Toki.
\newblock {Octet baryon masses in next-to-next-to-next-to-leading order
  covariant baryon chiral perturbation theory}.
\newblock {\em JHEP}, 12:073, 2012.
\newblock \url{https://doi.org/10.1007/JHEP12(2012)073}.

\bibitem{chen2013}
Y.-H. Chen, D.-L. Yao, and H.~Q. Zheng.
\newblock {Analyses of pion-nucleon elastic scattering amplitudes up to
  $\mathcal{O}(p^4)$ in extended-on-mass-shell subtraction scheme}.
\newblock {\em Phys. Rev. D}, 87:054019, 2013.
\newblock \url{https://doi.org/10.1103/PhysRevD.87.054019}.

\bibitem{yao2016}
D.-L. Yao, D.~Siemens, V.~Bernard, E.~Epelbaum, A.~M. Gasparyan, J.~Gegelia,
  H.~Krebs, and U.-G. Mei{\ss}ner.
\newblock {Pion-nucleon scattering in covariant baryon chiral perturbation
  theory with explicit Delta resonances}.
\newblock {\em JHEP}, 05:038, 2016.
\newblock \url{https://doi.org/10.1007/JHEP05(2016)038}.

\bibitem{lu2022}
J.-X. Lu, C.-X. Wang, Y.~Xiao, L.-S. Geng, J.~Meng, and P.~Ring.
\newblock {Accurate Relativistic Chiral Nucleon-Nucleon Interaction up to
  Next-to-Next-to-Leading Order}.
\newblock {\em Phys. Rev. Lett.}, 128:142002, 2022.
\newblock \url{https://doi.org/10.1103/PhysRevLett.128.142002}.

\bibitem{wise1992}
M.~B. Wise.
\newblock {Chiral perturbation theory for hadrons containing a heavy quark}.
\newblock {\em Phys. Rev. D}, 45:R2188, 1992.
\newblock \url{https://doi.org/10.1103/PhysRevD.45.R2188}.

\bibitem{wang2019}
B.~Wang, L.~Meng, and S.-L. Zhu.
\newblock {Hidden-charm and hidden-bottom molecular pentaquarks in chiral
  effective field theory}.
\newblock {\em JHEP}, 11:108, 2019.
\newblock \url{https://doi.org/10.1007/JHEP11(2019)108}.

\bibitem{meng2020}
L.~Meng, B.~Wang, and S.-L. Zhu.
\newblock {$\Sigma_cN$ interaction in chiral effective field theory}.
\newblock {\em Phys. Rev. C}, 101:064002, 2020.
\newblock \url{https://doi.org/10.1103/PhysRevC.101.064002}.

\bibitem{wang2020}
B.~Wang, L.~Meng, and S.-L. Zhu.
\newblock {$D^{(\ast)}N$ interaction and the structure of $\Sigma_c(2800)$ and
  $\Lambda_c(2940)$ in chiral effective field theory}.
\newblock {\em Phys. Rev. D}, 101:094035, 2020.
\newblock \url{https://doi.org/10.1103/PhysRevD.101.094035}.

\bibitem{chen2022}
K.~Chen, B.-L. Huang, B.~Wang, and S.-L. Zhu.
\newblock {$\Sigma_c\Sigma_c$ interactions in chiral effective field theory}.
\newblock 2022.
\newblock {Preprint at \url{https://arxiv.org/abs/2204.13316}}.

\bibitem{meng2022}
L.~Meng, B.~Wang, G.-J. Wang, and S.-L. Zhu.
\newblock Chiral perturbation theory for heavy hadrons and chiral effective
  field theory for heavy hadronic molecules.
\newblock 2022.
\newblock {Preprint at \url{https://arxiv.org/abs/2204.08716}}.

\bibitem{aube2003}
B.~Aubert et~al.
\newblock {Observation of a narrow meson decaying to $D_s^+\pi^0$ at a mass of
  2.32-GeV/c$^2$}.
\newblock {\em Phys. Rev. Lett.}, 90:242001, 2003.
\newblock \url{https://doi.org/10.1103/PhysRevLett.90.242001}.

\bibitem{krok2003}
P.~Krokovny et~al.
\newblock {Observation of the ${D}_{sJ}(2317)$ and ${D}_{sJ}(2457)$ in $B$
  Decays}.
\newblock {\em Phys. Rev. Lett.}, 91:262002, 2003.
\newblock \url{https://doi.org/10.1103/PhysRevLett.91.262002}.

\bibitem{bess2003}
D.~Besson et~al.
\newblock {Observation of a narrow resonance of mass $2.46$ GeV${/c}^{2}$
  decaying to ${D}_{s}^{*+}{\ensuremath{\pi}}^{0}$ and confirmation of the
  ${D}_{\mathrm{sJ}}^{*}(2317)$ state}.
\newblock {\em Phys. Rev. D}, 68:032002, 2003.
\newblock \url{https://doi.org/10.1103/PhysRevD.68.032002}.

\bibitem{choi2003}
S.~K. Choi et~al.
\newblock {Observation of a narrow charmonium-like state in exclusive $B^\pm
  \to K^\pm \pi^+ \pi^- J/\psi$ decays}.
\newblock {\em Phys. Rev. Lett.}, 91:262001, 2003.
\newblock \url{https://doi.org/10.1103/PhysRevLett.91.262001}.

\bibitem{liux2013}
X.~Liu.
\newblock {An overview of $XYZ$ new particles}.
\newblock {\em Chin. Sci. Bull.}, 59:3815--3830, 2014.
\newblock \url{https://doi.org/10.1007/s11434-014-0407-2}.

\bibitem{xie2017}
J.-J. Xie, W.-H. Liang, and E.~Oset.
\newblock {Hidden charm pentaquark and $\Lambda(1405)$ in the $\Lambda^0_b \to
  \eta_c K^- p (\pi \Sigma)$ reaction}.
\newblock {\em Phys. Lett. B}, 777:447--452, 2018.
\newblock \url{https://doi.org/10.1016/j.physletb.2017.12.064}.

\bibitem{ma2018}
Y.-L. Ma and M.~Harada.
\newblock {Chiral partner structure of doubly heavy baryons with heavy quark
  spin-flavor symmetry}.
\newblock {\em J. Phys. G}, 45:075006, 2018.
\newblock \url{https://doi.org/10.1088/1361-6471/aac86e}.

\bibitem{liu2019}
Y.-R. Liu, H.-X. Chen, W.~Chen, X.~Liu, and S.-L. Zhu.
\newblock {Pentaquark and Tetraquark states}.
\newblock {\em Prog. Part. Nucl. Phys.}, 107:237--320, 2019.
\newblock \url{https://doi.org/10.1016/j.ppnp.2019.04.003}.

\bibitem{lian2020}
W.-H. Liang, N.~Ikeno, and E.~Oset.
\newblock {$\Upsilon(nl)$ decay into $ B^{(*)} \bar B^{(*)}$}.
\newblock {\em Phys. Lett. B}, 803:135340, 2020.
\newblock \url{https://doi.org/10.1016/j.physletb.2020.135340}.

\bibitem{dong2020}
Y.~Dong, P.~Shen, F.~Huang, and Z.~Zhang.
\newblock {Selected strong decays of pentaquark State $P_c(4312)$ in a chiral
  constituent quark model}.
\newblock {\em Eur. Phys. J. C}, 80:341, 2020.
\newblock \url{https://doi.org/10.1140/epjc/s10052-020-7890-1}.

\bibitem{wu2021}
Qi~Wu, Dian-Yong Chen, Wen-Hua Qin, and Gang Li.
\newblock {Production of $Z_{cs}$ in B and $B_s$ decays}.
\newblock {\em Eur. Phys. J. C}, 82(6):520, 2022.
\newblock \url{https://doi.org/10.1140/epjc/s10052-022-10465-z}.

\bibitem{deng2022}
C.~Deng and S.-L. Zhu.
\newblock {$T_{cc}^{+}$ and its partners}.
\newblock {\em Phys. Rev. D}, 105:054015, 2022.
\newblock \url{https://doi.org/10.1103/PhysRevD.105.054015}.

\bibitem{deng20221}
Cheng-Rong Deng and Shi-Lin Zhu.
\newblock {Decoding the double heavy tetraquark state $T^+_{cc}$}.
\newblock {\em Sci. Bull.}, 67:1522, 2022.
\newblock \url{https://doi.org/10.1016/j.scib.2022.06.016}.

\bibitem{he2022}
J.~He and X.~Liu.
\newblock {The quasi-fission phenomenon of double charm $T_{cc}^+$ induced by
  nucleon}.
\newblock {\em Eur. Phys. J. C}, 82:387, 2022.
\newblock \url{https://doi.org/10.1140/epjc/s10052-022-10363-4}.

\bibitem{chenhx2022}
Hua-Xing Chen, Wei Chen, Xiang Liu, Yan-Rui Liu, and Shi-Lin Zhu.
\newblock {An updated review of the new hadron states}.
\newblock {\em Rept. Prog. Phys.}, 86(2):026201, 2023.
\newblock \url{https://doi.org/10.1088/1361-6633/aca3b6}.

\bibitem{wang2022}
Zhi-Hui Wang and Guo-Li Wang.
\newblock {Two-body strong decays of the 2P and 3P charmonium states}.
\newblock {\em Phys. Rev. D}, 106(5):054037, 2022.
\newblock \url{https://doi.org/10.1103/PhysRevD.106.054037}.

\bibitem{dai2022}
L.~R. Dai, R.~Molina, and E.~Oset.
\newblock {Looking for the exotic X0(2866) and its JP=1+ partner in the
  B\textasciimacron{}0\textrightarrow{}D(*)+K-K(*)0 reactions}.
\newblock {\em Phys. Rev. D}, 105(9):096022, 2022.
\newblock \url{https://doi.org/10.1103/PhysRevD.105.096022}.

\bibitem{barn2003}
T.~Barnes, F.~E. Close, and H.~J. Lipkin.
\newblock Implications of a {DK} molecule at 2.32 {G}e{V}.
\newblock {\em Phys. Rev. D}, 68:054006, 2003.
\newblock \url{https://doi.org/10.1103/PhysRevD.68.054006}.

\bibitem{beve2003}
E.~van Beveren and G.~Rupp.
\newblock Observed ${D}_{s}(2317)$ and tentative $d(2100--2300)$ as the charmed
  cousins of the light scalar nonet.
\newblock {\em Phys. Rev. Lett.}, 91:012003, 2003.
\newblock \url{https://doi.org/10.1103/PhysRevLett.91.012003}.

\bibitem{bali2003}
G.~S. Bali.
\newblock ${D}_{\mathrm{sj}}^{+}(2317):$ what can the lattice say?
\newblock {\em Phys. Rev. D}, 68:071501, 2003.
\newblock \url{https://doi.org/10.1103/PhysRevD.68.071501}.

\bibitem{dmit2005}
V.~Dmitra\ifmmode \check{s}\else \v{s}\fi{}inovi\ifmmode~\acute{c}\else
  \'{c}\fi{}.
\newblock ${D}_{s0}^{+}(2317)$-${D}_{0}(2308)$ mass difference as evidence for
  tetraquarks.
\newblock {\em Phys. Rev. Lett.}, 94:162002, 2005.
\newblock \url{https://doi.org/10.1103/PhysRevLett.94.162002}.

\bibitem{guo2006}
F.-K. Guo, P.-N. Shen, H.-C. Chiang, R.-G. Ping, and B.-S. Zou.
\newblock {Dynamically generated 0+ heavy mesons in a heavy chiral unitary
  approach}.
\newblock {\em Phys. Lett. B}, 641:278--285, 2006.
\newblock \url{https://doi.org/10.1016/j.physletb.2006.08.064}.

\bibitem{guo2007}
F.-K. Guo, P.-N. Shen, and H.-C. Chiang.
\newblock Dynamically generated 1+ heavy mesons.
\newblock {\em Phys. Lett. B}, 647:133--139, 2007.
\newblock \url{https://doi.org/10.1016/j.physletb.2007.01.050}.

\bibitem{flyn2007}
J.~M. Flynn and J.~Nieves.
\newblock {Elastic s-wave $B\pi$, $D\pi$, $D K$ and $K \pi$ scattering from
  lattice calculations of scalar form-factors in semileptonic decays}.
\newblock {\em Phys. Rev. D}, 75:074024, 2007.
\newblock \url{https://doi.org/10.1103/PhysRevD.75.074024}.

\bibitem{lutz2008}
Matthias F.~M. Lutz and Madeleine Soyeur.
\newblock {Radiative and isospin-violating decays of D(s)-mesons in the
  hadrogenesis conjecture}.
\newblock {\em Nucl. Phys. A}, 813:14--95, 2008.
\newblock \url{https://doi.org/10.1016/j.nuclphysa.2008.09.003}.

\bibitem{mohl2013}
D.~Mohler, C.~B. Lang, L.~Leskovec, S.~Prelovsek, and R.~M. Woloshyn.
\newblock ${D}_{s0}^{*}\mathbf{(}2317\mathbf{)}$ meson and {D}-meson-kaon
  scattering from lattice {QCD}.
\newblock {\em Phys. Rev. Lett.}, 111:222001, 2013.
\newblock \url{https://doi.org/10.1103/PhysRevLett.111.222001}.

\bibitem{liu2013}
L.~Liu, K.~Orginos, F.-K. Guo, C.~Hanhart, and U.-G. Mei\ss{}ner.
\newblock Interactions of charmed mesons with light pseudoscalar mesons from
  lattice {QCD} and implications on the nature of the ${D}_{s0}^{*}(2317)$.
\newblock {\em Phys. Rev. D}, 87:014508, 2013.
\newblock \url{https://doi.org/10.1103/PhysRevD.87.014508}.

\bibitem{alex2020}
C.~Alexandrou, J.~Berlin, J.~Finkenrath, T.~Leontiou, and M.~Wagner.
\newblock {Tetraquark interpolating fields in a lattice QCD investigation of
  the $D_{s0}^\ast(2317)$ meson}.
\newblock {\em Phys. Rev. D}, 101:034502, 2020.
\newblock \url{https://doi.org/10.1103/PhysRevD.101.034502}.

\bibitem{tan2021}
Y.~Tan and J.~Ping.
\newblock {$D^*_{s0}(2317)$ and $D_{s1}(2460)$ in an unquenched quark model}.
\newblock 2021.
\newblock {Preprint at \url{https://arxiv.org/abs/2111.04677}}.

\bibitem{cheu2021}
G.~K.~C. Cheung, C.~E. Thomas, D.~J. Wilson, G.~Moir, M.~Peardon, and S.~Ryan.
\newblock {DK I = 0, $D\overline{K}$ I = 0, 1 scattering and the $
  {D}_{s0}^{\ast } $(2317) from lattice QCD}.
\newblock {\em JHEP}, 02:100, 2021.
\newblock \url{https://doi.org/10.1007/JHEP02(2021)100}.

\bibitem{yang2022}
Z.~Yang, G.-J. Wang, J.-J. Wu, M.~Oka, and S.-L. Zhu.
\newblock {Novel Coupled Channel Framework Connecting the Quark Model and
  Lattice QCD for the Near-threshold Ds States}.
\newblock {\em Phys. Rev. Lett.}, 128:112001, 2022.
\newblock \url{https://doi.org/10.1103/PhysRevLett.128.112001}.

\bibitem{chen2017}
H.-X. Chen, W.~Chen, X.~Liu, Y.-R. Liu, and S.-L. Zhu.
\newblock {A review of the open charm and open bottom systems}.
\newblock {\em Rept. Prog. Phys.}, 80:076201, 2017.
\newblock \url{https://doi.org/10.1088/1361-6633/aa6420}.

\bibitem{liu2009}
Y.-R. Liu, X.~Liu, and S.-L. Zhu.
\newblock {Light Pseudoscalar Meson and Heavy Meson Scattering Lengths}.
\newblock {\em Phys. Rev. D}, 79:094026, 2009.
\newblock \url{https://doi.org/10.1103/PhysRevD.79.094026}.

\bibitem{huan2022}
B.-L. Huang, Z.-Y. Lin, and S.-L. Zhu.
\newblock {Light pseudoscalar meson and heavy meson scattering lengths to
  $\mathcal{O}(p^4)$ in heavy meson chiral perturbation theory}.
\newblock {\em Phys. Rev. D}, 105:036016, 2022.
\newblock \url{https://doi.org/10.1103/PhysRevD.105.036016}.

\bibitem{lutz2004}
E.~E. Kolomeitsev and M.~F.~M. Lutz.
\newblock {On heavy light meson resonances and chiral symmetry}.
\newblock {\em Phys. Lett. B}, 582:39--48, 2004.
\newblock \url{https://doi.org/10.1016/j.physletb.2003.10.118}.

\bibitem{guo2009}
F.-K. Guo, C.~Hanhart, and U.-G. Mei{\ss}ner.
\newblock {Interactions between heavy mesons and Goldstone bosons from chiral
  dynamics}.
\newblock {\em Eur. Phys. J. A}, 40:171--179, 2009.
\newblock \url{https://doi.org/10.1140/epja/i2009-10762-1}.

\bibitem{geng2010}
L.~S. Geng, N.~Kaiser, J.~Martin-Camalich, and W.~Weise.
\newblock Low-energy interactions of nambu-goldstone bosons with $d$ mesons in
  covariant chiral perturbation theory.
\newblock {\em Phys. Rev. D}, 82:054022, 2010.
\newblock \url{https://doi.org/10.1103/PhysRevD.82.054022}.

\bibitem{wang2012}
P.~Wang and X.~G. Wang.
\newblock Publisher's note: Study of ${0}^{\mathbf{+}}$ states with open charm
  in the unitarized heavy meson chiral approach [phys. rev. d 86, 014030
  (2012)].
\newblock {\em Phys. Rev. D}, 86:039903, Aug 2012.
\newblock \url{https://doi.org/10.1103/PhysRevD.86.039903}.

\bibitem{alte2014}
M.~Altenbuchinger, L.-S. Geng, and W.~Weise.
\newblock {Scattering lengths of Nambu-Goldstone bosons off $D$ mesons and
  dynamically generated heavy-light mesons}.
\newblock {\em Phys. Rev. D}, 89:014026, 2014.
\newblock \url{https://doi.org/10.1103/PhysRevD.89.014026}.

\bibitem{yao2015}
D.-L. Yao, M.-L. Du, F.-K. Guo, and U.-G. Mei\ss{}ner.
\newblock {One-loop analysis of the interactions between charmed mesons and
  Goldstone bosons}.
\newblock {\em JHEP}, 11:058, 2015.
\newblock \url{https://doi.org/10.1007/JHEP11(2015)058}.

\bibitem{guo2019}
Z.-H. Guo, L.~Liu, U.-G. Mei\ss{}ner, J.~A. Oller, and A.~Rusetsky.
\newblock {Towards a precise determination of the scattering amplitudes of the
  charmed and light-flavor pseudoscalar mesons}.
\newblock {\em Eur. Phys. J. C}, 79:13, 2019.
\newblock \url{https://doi.org/10.1140/epjc/s10052-018-6518-1}.

\bibitem{bora1997}
B.~Borasoy and U.-G. Mei{\ss}ner.
\newblock {Chiral expansion of baryon masses and sigma-terms}.
\newblock {\em Annals Phys.}, 254:192--232, 1997.
\newblock \url{https://doi.org/10.1006/aphy.1996.5630}.

\bibitem{gass1991}
J.~Gasser and U.-G. Mei{\ss}ner.
\newblock {On the phase of epsilon-prime}.
\newblock {\em Phys. Lett. B}, 258:219--224, 1991.
\newblock \url{https://doi.org/10.1016/0370-2693(91)91235-N}.

\bibitem{eric1988}
T.~E.~O. Ericson and W.~Weise.
\newblock {\em {Pions and nuclei}}.
\newblock Clarendon Press, Oxford, UK, 1988.

\bibitem{moir2016}
G.~Moir, M.~Peardon, S.~Ryan, C.~E. Thomas, and D.~J. Wilson.
\newblock {Coupled-Channel $D\pi$, $D\eta$ and $D_{s}\bar{K}$ Scattering from
  Lattice QCD}.
\newblock {\em JHEP}, 10:011, 2016.
\newblock \url{https://doi.org/10.1007/JHEP10(2016)011}.

\bibitem{walk2009}
A.~Walker-Loud et~al.
\newblock Light hadron spectroscopy using domain wall valence quarks on an
  asqtad sea.
\newblock {\em Phys. Rev. D}, 79:054502, 2009.
\newblock \url{https://doi.org/10.1103/PhysRevD.79.054502}.

\bibitem{doba2014}
J.~Dobaczewski, W.~Nazarewicz, and P.-G. Reinhard.
\newblock {Error estimates of theoretical models: a Guide}.
\newblock {\em J. Phys. G}, 41:074001, 2014.
\newblock \url{https://doi.org/10.1088/0954-3899/41/7/074001}.

\bibitem{carl2016}
Carlsson~B. D. et~al.
\newblock {Uncertainty analysis and order-by-order optimization of chiral
  nuclear interactions}.
\newblock {\em Phys. Rev. X}, 6:011019, 2016.
\newblock \url{https://doi.org/10.1103/PhysRevX.6.011019}.

\bibitem{chan2018}
C.~C.~Chang $et$ $al.$.
\newblock {A per-cent-level determination of the nucleon axial coupling from
  quantum chromodynamics}.
\newblock {\em Nature}, 558:91--94, 2018.
\newblock \url{https://doi.org/10.1038/s41586-018-0161-8}.

\bibitem{mark2019}
B.~M{\"a}rkisch $et$ $al.$.
\newblock {Measurement of the weak axial-vector coupling constant in the decay
  of free neutrons using a pulsed cold neutron beam}.
\newblock {\em Phys. Rev. Lett.}, 122:242501, 2019.
\newblock \url{https://doi.org/10.1103/PhysRevLett.122.242501}.

\bibitem{lee1994}
Chang-Hwan Lee, Hong Jung, Dong-Pil Min, and Mannque Rho.
\newblock {Kaon - nucleon scattering from chiral Lagrangians}.
\newblock {\em Phys. Lett. B}, 326:14--20, 1994.
\newblock \url{https://doi.org/10.1016/0370-2693(94)91185-1}.

\bibitem{pdg2020}
P.~A. Zyla et~al.
\newblock {Review of Particle Physics}.
\newblock {\em Prog.Theor.Exp.Phys.}, 2020:083C01, 2020.
\newblock \url{https://doi.org/10.1093/ptep/ptaa104}.

\bibitem{guo2018}
Xiao-Yu Guo, Yonggoo Heo, and Matthias F.~M. Lutz.
\newblock {On chiral extrapolations of charmed meson masses and coupled-channel
  reaction dynamics}.
\newblock {\em Phys. Rev. D}, 98:014510, 2018.
\newblock \url{https://doi.org/10.1103/PhysRevD.98.014510}.

\bibitem{guo2022}
Xiao-Yu Guo, Yonggoo Heo, and Matthias F.~M. Lutz.
\newblock {From lattice QCD to predictions of scattering phase shifts at the
  physical point}.
\newblock {\em PoS}, LATTICE2021:601, 2022.
\newblock \url{https://doi.org/10.22323/1.396.0601}.

\bibitem{greg2021}
E.~B. Gregory, F.-K. Guo, C.~Hanhart, S.~Krieg, and T.~Luu.
\newblock Confirmation of the existence of an exotic state in the $\pi{D}$
  system.
\newblock 2021.
\newblock {Preprint at \url{https://arxiv.org/abs/2106.15391}}.

\bibitem{alba2017}
M.~Albaladejo, P.~Fernandez-Soler, F.-K. Guo, and J.~Nieves.
\newblock {Two-pole structure of the $D^\ast_0(2400)$}.
\newblock {\em Phys. Lett. B}, 767:465--469, 2017.
\newblock \url{https://doi.org/10.1016/j.physletb.2017.02.036}.

\end{thebibliography}

\end{document}